\documentclass{article}

\usepackage{microtype}
\usepackage{graphicx}
\usepackage{booktabs}
\usepackage{tabularx}
\usepackage{hyperref}
\usepackage[preprint]{icml2026}
\usepackage{amsmath}
\usepackage{amssymb}
\usepackage[font=small,labelfont=bf]{caption}
\usepackage{xcolor}
\IfFileExists{todonotes.sty}{\usepackage{todonotes}}{}
\usepackage{comment}
\usepackage[capitalize,noabbrev]{cleveref}
\crefname{appendix}{Appendix}{Appendices}
\DeclareMathOperator{\argmax}{arg\,max}

\newcommand{\Dnorm}{\left\lVert D \right\rVert_{2}}

\begin{document}

\twocolumn[

\icmltitle{Geometric Signatures of Conceptual Reorganization: A Counterfactual Embedding Framework for Detecting Scientific Revolutions}

\icmltitlerunning{Geometric Signatures of Conceptual Reorganization}

\begin{icmlauthorlist}
\icmlauthor{Dimitris Ntounis}{stanfordPH,slac}
\icmlauthor{Ariel Schwartzman}{slac}
\icmlauthor{Chris Chafe}{stanfordMUS}
\icmlauthor{Thomas A. Ryckman}{stanfordPHIL}
\end{icmlauthorlist}
\icmlaffiliation{stanfordPH}{Department of Physics, Stanford University, Stanford, CA, USA}
\icmlaffiliation{stanfordMUS}{Department of Music, Stanford University, Stanford, CA, USA}
\icmlaffiliation{stanfordPHIL}{
Department of Philosophy, Stanford University, Stanford, CA, USA}
\icmlaffiliation{slac}{SLAC National Accelerator Laboratory, Menlo Park, CA, USA}

\icmlcorrespondingauthor{Dimitris Ntounis}{dntounis@stanford.edu}

\vskip 0.3in
]
\printAffiliationsAndNotice{}

\begin{abstract}

We introduce document embedding geometry as a quantitative observable of conceptual reorganization and develop a counterfactual ablation framework for measuring how individual concepts influence the organization of scientific knowledge, providing a quantitative framework for detecting scientific revolutions. The observable is defined by the geometric perturbation induced when removing documents associated with a candidate concept from the embedding space before and after its historical emergence. Statistical validation is performed using five historical case studies spanning physics, mathematics, and machine learning: special relativity, Gödel's incompleteness theorems, the Higgs mechanism, deep learning, and the attention mechanism underlying transformer architectures. Across the historical case studies, the framework identifies measurable geometric signatures associated with conceptual reorganization, while the validation studies expose important limitations arising from document assignment and sparse historical data. These results establish embedding geometry as a medium for quantifying conceptual reorganization, providing a new approach for studying how scientific fields restructure over time.\footnote{Code and accompanying material are available at \url{https://github.com/Mapping-Innovation-Lab/geometric-signatures} and \url{https://mapping-innovation-lab.github.io/geometric-signatures-companion-website/}.}
\end{abstract}

\section{Introduction}

Understanding how knowledge is organized and how that organization changes over time has long been a central objective of the history and philosophy of science and, more recently, of computational approaches to understanding scientific change and innovation. Scientific knowledge is organized through a network of interconnected concepts whose relationships continually evolve as new discoveries emerge. While much of this evolution is gradual, periods of rapid conceptual reorganization occasionally reshape entire disciplines, fundamentally changing how previously established ideas are understood and connected. Such scientific advances do more than extend existing knowledge: they reorganize established understanding and redirect the trajectory of a field~\cite{kuhn1962structure}. Special relativity, Gödel’s incompleteness theorems, the Higgs mechanism, and, more recently, deep learning are prominent examples. Understanding these reorganizations has traditionally relied on historical analysis and qualitative interpretation. A quantitative description of how conceptual organization itself changes remains comparatively underdeveloped.

Existing quantitative approaches~\citep{fortunato2018science} have sought to characterize scientific change through citation networks, bibliometric indicators, topic evolution, semantic change, and other data-driven analyses. These methods have revealed important patterns of scientific development, including disruption, interdisciplinarity, and knowledge diffusion. However, they primarily characterize relationships external to the conceptual organization itself, such as citation structure, publication dynamics, or changes in topic prevalence, and often inherit the bias of external metadata.  In this work, we investigate a complementary question: \emph{can quantifiable conceptual reorganizations, including those associated with scientific revolutions, be detected through changes in the geometry of document embedding spaces?}

Modern embedding models represent documents as vectors in high-dimensional spaces whose geometry reflects statistical regularities learned directly from text. Since a document's position in embedding space encodes its semantic content~\citep{ethayarajh2019contextual}, these representations have become fundamental tools for information retrieval, clustering, recommendation, and scientific search. More fundamentally, embedding spaces may be viewed as learned \emph{semantic geometries}, in which the relative positions of concepts reflect the statistical structure of the scientific literature. Our starting point is the observation that scientific publications constitute the primary written record through which the conceptual organization of a field is communicated, refined, and extended. If embedding models capture sufficiently rich representations of this literature, then changes in conceptual organization should become observable through changes in the embedding geometry.

 Embedding models generally place documents discussing related concepts in nearby regions of the embedding space, while conceptually distinct documents tend to be more widely separated. Consequently, major conceptual reorganizations are expected to manifest not merely through the appearance of new documents, but through measurable changes in the global organization of the embedding space. Our central hypothesis is that concepts responsible for major conceptual reorganizations occupy distinctive structural roles within embedding geometry and that these roles can be measured quantitatively. This perspective suggests a shift in how embedding spaces are used. Rather than treating them solely as representations for retrieval or semantic similarity, we investigate whether their geometry can serve as a measurable scientific observable. In this view, conceptual reorganization is inferred not from the appearance of individual documents, but from the geometric response of the embedding space to controlled counterfactual perturbations.

The central conceptual contribution of this work is to promote embedding geometry from a descriptive semantic representation to a quantitative scientific observable that can be measured, perturbed, statistically validated, and compared across independent historical case studies. For this reason, we introduce a counterfactual ablation framework. For a candidate concept, we remove the documents associated with that concept, reconstruct the embedding geometry, and compare the resulting perturbation before and after the historical emergence of the concept. The underlying intuition is straightforward: if a concept fundamentally reorganizes an existing body of knowledge, removing it should perturb the post-emergence geometry substantially more than the pre-emergence geometry. The resulting asymmetry provides a quantitative observable of conceptual reorganization without relying on citation graphs, expert annotations, or supervised training.

We evaluate the proposed framework on five historical case studies spanning physics, mathematics, and machine learning: special relativity~\citep{einstein1905electrodynamics}, G\"odel's incompleteness theorems~\citep{godel1931undecidable}, the Higgs mechanism~\citep{higgs1964broken}, the deep learning revolution~\citep{krizhevsky2012imagenet}, and the attention mechanism~\citep{vaswani2017attention} underlying transformer architectures. These case studies were selected to span different scientific disciplines and historical contexts, enabling us to study the temporal evolution of conceptual organization under diverse conditions. Rather than attempting to discover previously unknown historical events, our objective is to determine whether embedding geometry captures measurable signatures consistent with historically recognized episodes of conceptual reorganization.

Specifically, this work makes three main contributions. First, we propose embedding geometry as a quantitative observable for studying conceptual reorganization. Second, we introduce a counterfactual ablation framework that measures the structural influence of individual concepts on embedding geometry. Third, we apply the approach across five case studies spanning physics, mathematics, and machine learning to study the temporal evolution of conceptual organization.

Our broader aim is to understand what the embedding spaces of modern language models can reveal about the organization and evolution of knowledge. Rather than replacing existing computational approaches to scientific change, we propose a complementary geometric perspective that treats embedding geometry as a quantitative observable of conceptual organization. Although the present work focuses on scientific and mathematical literature, the framework is naturally applicable to other text-rich domains in which conceptual structures evolve over time, opening opportunities for quantitatively studying the dynamics of ideas across disciplines.

The present work addresses the first step of this broader research program: establishing that historically recognized conceptual reorganizations leave measurable geometric signatures in embedding space. A long-term ambition is to develop quantitative models of knowledge evolution that may ultimately contribute to AI systems capable of assisting scientific discovery by identifying emerging conceptual reorganizations.

\section{Method}
\label{sec:method}

\begin{figure*}[t]
\centering
\includegraphics[width=0.95\textwidth]{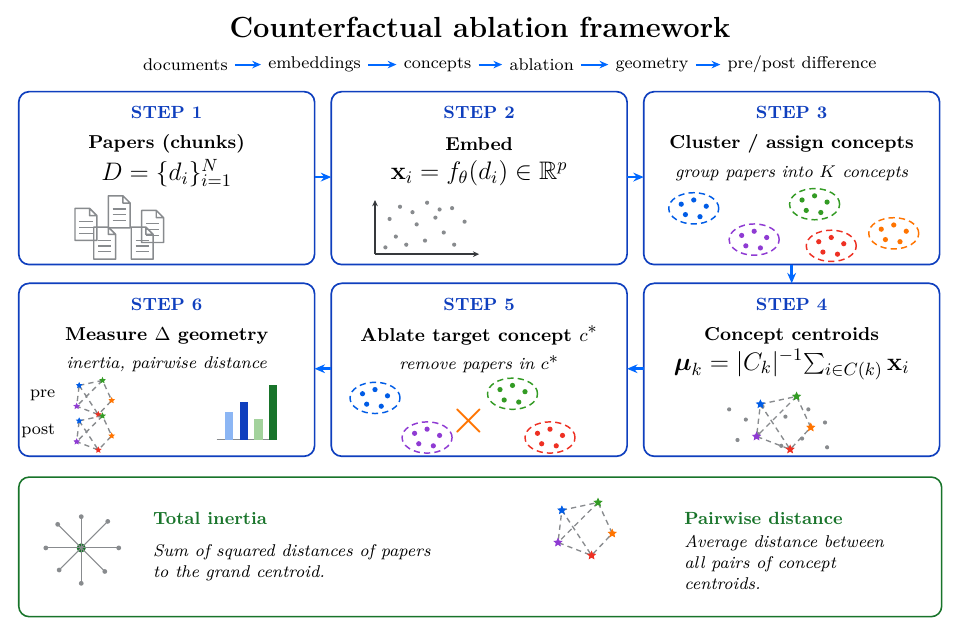}
\caption{Schematic of the proposed counterfactual ablation framework.}
\label{fig:overview}
\end{figure*}

The complete six-step framework is illustrated in \cref{fig:overview}. The first four steps construct the geometric representation described below. The final two steps implement the counterfactual framework by measuring the geometric perturbation induced by removing the documents associated with a target concept and comparing the resulting perturbation before and after a candidate pivot year.

\subsection{Representing conceptual organization}
\label{sec:embeddings}

Our framework represents both scientific documents and conceptual descriptions within a common embedding space. The representation consists of three steps. First, each scientific document is mapped to an embedding vector that captures its semantic content. Second, a small set of historically meaningful concepts is represented by embedding-based concept anchors. Finally, documents are associated with concepts through semantic similarity. This representation provides the geometric foundation for the counterfactual framework introduced in the next subsection.

Scientific documents are embedded using modern sentence-transformer models. Because many papers exceed the context window of these models, each document is divided into overlapping segments (chunks) of at most 300 words with a 40-word overlap. Each chunk is embedded independently, and the document representation is the $\ell_2$-normalized mean of its chunk embeddings, that is, the document's coordinate vector in the embedding space. This procedure incorporates information from throughout the document rather than only its opening text.

Rather than discovering concepts through unsupervised clustering, we define for each historical case a fixed set of $N=10$~\footnote{The choice of $N=10$ was established through a robustness study over $N\in\{5,8,10\}$, where the target concept attains rank one in four of the five cases at $N=10$.}  concepts consisting of the target concept together with nine contextual concepts spanning the principal subfields surrounding the historical transition. Each concept is represented by a short natural-language description. This design choice keeps the resulting geometry directly interpretable across historical case studies and embedding models, while isolating the structural role of historically identifiable concepts rather than latent statistical topics

Documents are assigned to concepts by comparing each document embedding with an ensemble of concept anchors. For each concept, five paraphrases of its natural-language description are embedded using the same encoder employed for the document corpus, and the concept anchor is the $\ell_2$-normalized mean of the five resulting description embeddings. Averaging across paraphrases reduces sensitivity to the wording of any individual description. Each document is assigned to the concept with the highest cosine similarity provided that two conditions are satisfied: the similarity exceeds a per-concept baseline threshold, and the margin to the second-best concept (\cref{eq:margin}) exceeds a confusion-aware threshold. Documents failing either criterion are labeled ``no match'' and excluded from the geometric analysis. Details of the threshold construction are given in \cref{sec:threshold}.

Finally, because the proposed framework should reflect conceptual organization rather than properties of a particular embedding model, we evaluate the complete framework using five sentence-transformer models spanning four distinct training paradigms (\cref{tab:embedding_models_main}). Each model is applied independently from document embedding through statistical validation, allowing the robustness of the geometric signatures to be assessed across substantially different embedding representations. Results for all historical case studies are presented in~\cref{sec:modelrobustness}.

\begin{table*}[t]
\centering
\small
\caption{Embedding models used in this study. All results reported in the main body use e5-base; the remaining four models are used for the robustness study of \cref{sec:modelrobustness}.}
\label{tab:embedding_models_main}
\begin{tabular}{@{}llrl@{}}
\toprule
Short name & Training paradigm & Dim. & Reference \\
\midrule
e5-base   & Contrastive, multilingual & 768  & \citet{wang2024e5} \\
mxbai     & Contrastive               & 1024 & \citet{emde2024mxbai} \\
bge-m3    & Multi-task, multilingual   & 1024 & \citet{chen2024bge} \\
specter2  & Citation-prediction        & 768  & \citet{singh2023specter2} \\
MiniLM    & Distilled, English         & 384  & \citet{wang2020minilm,reimers2019sentencebert} \\
\bottomrule
\end{tabular}
\end{table*}

\subsection{Counterfactual framework}
\label{sec:ablation}

 We define the counterfactual framework for conceptual reorganization through the geometric perturbation induced by a \emph{counterfactual ablation}, in which all documents associated with a given concept are removed from the corpus and the resulting geometry is recomputed. The magnitude of this perturbation defines the standardized response used throughout the remainder of this paper. The framework is designed to quantify not the introduction of a concept itself, but its structural impact on the organization of the surrounding scientific literature.

For a given target concept, the counterfactual measurement proceeds as follows. First, all documents associated with the target concept are removed from the corpus. Second, concept centroids are recomputed within rolling windows of radius $r$ (spanning year $\pm r$), with $r=2$ for the three smaller corpora and $r=1$ for the two large machine-learning corpora; see \cref{tab:corpora}. For each concept $c$ present within a given window, the centroid $\mathbf{c}_c(t)$ is the $\ell_2$-normalized mean of the document embeddings assigned to that concept. Third, the geometry of the concept space is evaluated for each time window using the observables defined below. Fourth, the geometric perturbation induced by the ablation is computed by comparing the baseline and ablated geometries for every year. Finally, the perturbations before and after a candidate pivot year are compared statistically to quantify the asymmetry associated with the emergence of the concept.

Two complementary observables are used to characterize the geometry of the concept space. They quantify different aspects of conceptual organization: the overall spread of the concept space and the average separation between concepts. In the expressions below, $N(t)$ denotes the number of concepts with at least one assigned document in the rolling window centered on year $t$.

\paragraph{Total inertia ($d_I$).}

Let $\mathbf{c}_c(t)$ denote the centroid of concept $c$ within the rolling window centered on year $t$, and let

\[
\bar{\mathbf{c}}(t)=\frac{1}{N(t)}\sum_{c=1}^{N(t)}\mathbf{c}_c(t)
\]

be the corresponding average centroid. Total inertia is defined as

\begin{equation}
d_I(t)=
\sum_{c=1}^{N(t)}
\left\|
\mathbf{c}_c(t)-\bar{\mathbf{c}}(t)
\right\|^2.
\end{equation}

This quantity measures the overall spread of the concept space. Concepts that introduce genuinely new directions into the embedding geometry increase the total inertia disproportionately compared with small local drifts.

\paragraph{Mean pairwise cosine distance ($d_P$).}

The second observable is the average cosine distance between all concept pairs,

\begin{equation}
d_P(t)=
\binom{N(t)}{2}^{-1}
\sum_{i<j}
\left(
1-\cos(\mathbf{c}_i(t),\mathbf{c}_j(t))
\right).
\end{equation}

Unlike total inertia, this observable measures the average angular separation between concepts and is insensitive to vector magnitude, making it a natural measure of semantic separation in dense embedding spaces.

For each observable, the yearly geometric perturbation is defined as the difference between the baseline and ablated geometries,

\begin{equation}
\Delta(t)=g^{\rm base}(t)-g^{\rm abl}(t),
\end{equation}

where $g$ denotes either $d_I$ or $d_P$. Given a candidate pivot year $t^{*}$, the average perturbations before and after the pivot are

\begin{equation}
\bar{\Delta}_{\rm pre}
=
\frac{1}{n_{\rm pre}}
\sum_{t<t^*}\Delta(t),
\qquad
\bar{\Delta}_{\rm post}
=
\frac{1}{n_{\rm post}}
\sum_{t\ge t^*}\Delta(t),
\label{eq:delta_bar}
\end{equation}

where $n_{\rm pre}$ and $n_{\rm post}$ denote the number of yearly observations before and after the pivot.

The strength of the asymmetry is quantified using Cohen's $D$~\citep{cohen1988statistical},

\begin{align}
D
&=
\frac{\bar{\Delta}_{\rm post}
-
\bar{\Delta}_{\rm pre}}
{s_{\rm pool}},
\\
s_{\rm pool}^2
&=
\frac{
(n_{\rm pre}-1)s_{\rm pre}^2
+
(n_{\rm post}-1)s_{\rm post}^2
}
{
n_{\rm pre}
+
n_{\rm post}
-
2
},
\label{eq:cohens_d}
\end{align}

where $s_{\rm pre}$ and $s_{\rm post}$ are the sample standard deviations of the yearly perturbations before and after the pivot, respectively, and $s_{\rm pool}$ is their pooled standard deviation. Cohen's $D$ is a standardized geometric perturbation, allowing direct comparison between geometric observables with different numerical scales while avoiding dependence on their absolute normalization. Throughout the paper we denote this standardized response statistic by the uppercase symbol $D$ to distinguish it from the geometric observables $d_I$ and $d_P$.

Different conceptual innovations need not perturb every aspect of the embedding geometry equally. Some primarily introduce new geometric directions and therefore increase the overall spread of the concept space, while others predominantly reorganize relationships among existing concepts. We therefore evaluate both throughout the analysis. For concept ranking, we report the largest standardized response, $\max(|D_I|,|D_P|)$. For significance, however, collapsing the two
  observables to their maximum discards the second dimension, which carries
  information precisely in those diffuse transitions where neither metric dominates.
  We therefore assess significance jointly on the $(D_I,D_P)$ plane using a
  covariance-free co-dominance tail.\footnote{We use the maximum of the two standardized responses for ranking because different conceptual reorganizations perturb different aspects of embedding geometry. Ranking by the dominant response allows each case to be characterized by the observable on which it exhibits its strongest geometric signature, while statistical significance is assessed jointly using both observables.} Writing $\boldsymbol\mu=(\mu_I,\mu_P)$ for the
  mean of the null cloud and $(D_I^{\rm obs},D_P^{\rm obs})$ for the observation,
  \begin{equation}
  p_{\rm co} = \frac{1 + M}{N + 1},
  \label{eq:codominance}
  \end{equation}
  where $N$ is the number of null realizations and $M$ counts those that are at
  least as extreme as the observation on \emph{both} axes simultaneously,
  \begin{equation}
  \begin{aligned}
  M = \#\bigl\{ k :\ &|D_{I,k}-\mu_I| \ge |D_I^{\rm obs}-\mu_I| \\
     \text{and}\ &|D_{P,k}-\mu_P| \ge |D_P^{\rm obs}-\mu_P| \bigr\}.
  \end{aligned}
  \label{eq:codominance-count}
  \end{equation}
  We call a response statistically significant when its permutation-based $p$-value is below $0.05$.  We do not rank on the joint magnitude $\sqrt{D_I^2+D_P^2}$ because it is not covariance-free: the null correlation between $D_I$ and $D_P$ varies from $+0.11$ to $+1.00$ across the five cases, so a fixed contour in the plane corresponds to a different tail probability in each, and rankings built on it would not be comparable across cases. Ranking on a genuinely sign-aware statistic does not change the result. Ranking the ten concepts of each case by the co-dominance tail probability of their signed $(D_I, D_P)$ pair, each against its own permutation null of 5000 draws, reproduces the ranking of \cref{tab:crosscase} in four of the five cases and raises the attention mechanism from fourth to third. In the deep-learning case, where ranking on the joint magnitude would place a rival concept first, the sign-aware tail restores the target to first place, because that rival's two components carry opposite signs and therefore fall in a far less extreme region of its own null. The absolute value in $\max(|D_I|,|D_P|)$ therefore does not drive the reported ranking.

\subsection{Localizing conceptual reorganization}
\label{sec:grid}

\begin{figure*}
\centering
\includegraphics[width=\textwidth]{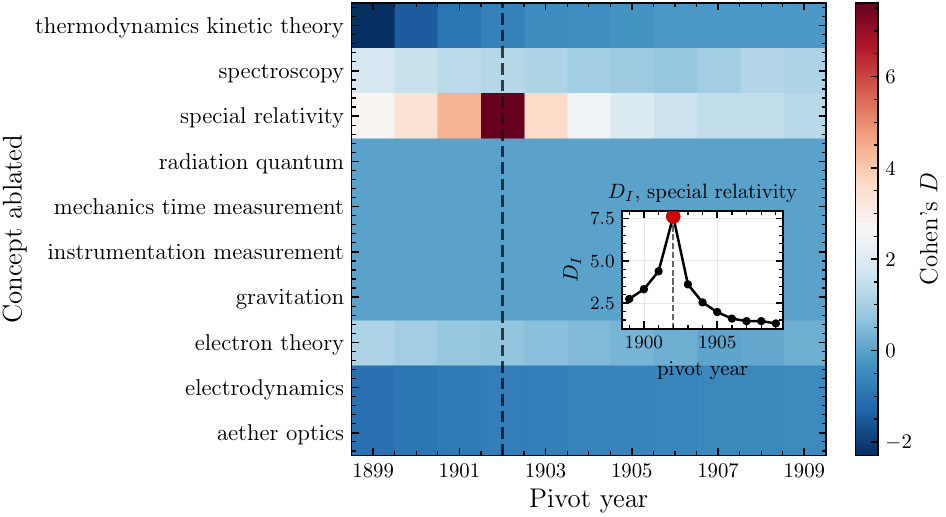}
\caption{Concept $\times$ pivot-year heatmap for special relativity.
         Each cell shows Cohen's $D_{I}$ for ablating the row's concept and splitting pre- vs.\ post-pivot at the column's year. The inset plots $D_I$ along the special-relativity row of the same grid.}
\label{fig:grid}
\end{figure*}

\begin{figure*}[t]
\centering
\includegraphics[width=0.85\textwidth]{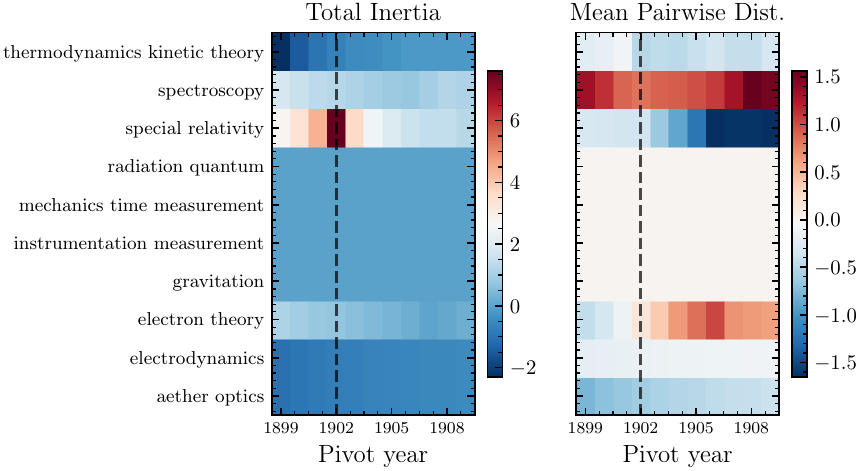}
\caption{Per-metric concept $\times$ pivot-year heatmaps for the special relativity case.}
\label{fig:grid3}
\end{figure*}

Conceptual reorganizations are expected to occur over finite historical periods rather than at predetermined dates. To localize these transitions in a data-driven manner, we evaluate $D$ over all combinations of candidate concepts and pivot years, producing a concept $\times$ pivot-year grid for each historical case. \Cref{fig:grid} illustrates the resulting grid for the special relativity corpus, with the corresponding per-metric decomposition into $D_I$ and $D_P$ shown in \cref{fig:grid3}.

Rather than fixing a historically motivated pivot year \emph{a priori}, this approach allows the data to determine when the geometric perturbation associated with a concept is strongest. Each candidate pivot year is a possible split between the pre- and post-pivot periods; the data-driven pivot $t^{*}$ is the candidate that maximizes the standardized response for the target concept. For each candidate pivot year, the counterfactual procedure described in \cref{sec:ablation} is applied independently, yielding a standardized response for every concept--pivot-year combination. The lower time bound of the scan is chosen so that every candidate pivot year has sufficient pre-pivot documents for reliable standardized-response estimation.

Each cell of the concept $\times$ pivot-year grid represents the standardized response obtained by applying the counterfactual framework to one concept using one candidate pivot year. The two observables carry different sign information. Because removing a concept can never increase the total inertia of the remaining centroids, $\Delta_I(t)\ge 0$ identically, so the sign of $D_I$ contrasts the magnitude of the pre- and post-pivot disruption: positive values mark a concept whose geometric footprint appears or grows at the pivot, and negative values one whose footprint was already largest before it. The pairwise-distance perturbation is instead signed, with a positive $\Delta_P$ identifying a concept sitting farther from its neighbors than they sit from one another, occupying a peripheral direction, while a negative $\Delta_P$ identifies a concept lying inside the ambient spread, so that its removal leaves the remaining concepts more widely separated. The sign of $D_P$ accordingly contrasts this relative position before and after the pivot rather than measuring the size of the disruption.

The concept $\times$ pivot-year grid may therefore be viewed as a two-dimensional map of the standardized response, describing how geometric perturbations depend jointly on conceptual identity and historical time. For each historical case, the pivot year, which corresponds to the time at which the target concept produces the strongest geometric perturbation, is defined as

\begin{equation}
t^{*}_{\rm case}
=
\arg\max_t
|D_{\rm case}(\mathrm{target},t)|.
\label{eq:pivot}
\end{equation}

Because a single metric dominates at each selected pivot, this $\ell_\infty$ choice coincides with the $\Dnorm$ grid-maximum for four of the five cases, with Gödel differing by two years. Moreover, the pivot is a localization step whose freedom is separately corrected by the look-elsewhere scan of \cref{sec:validation}, so per-case significance is never conditioned on it. The resulting pivot years for all five historical case studies are summarized in \cref{tab:crosscase}.

\subsection{Statistical validation}
\label{sec:validation}

\begin{figure*}
\centering
\includegraphics[width=\textwidth]{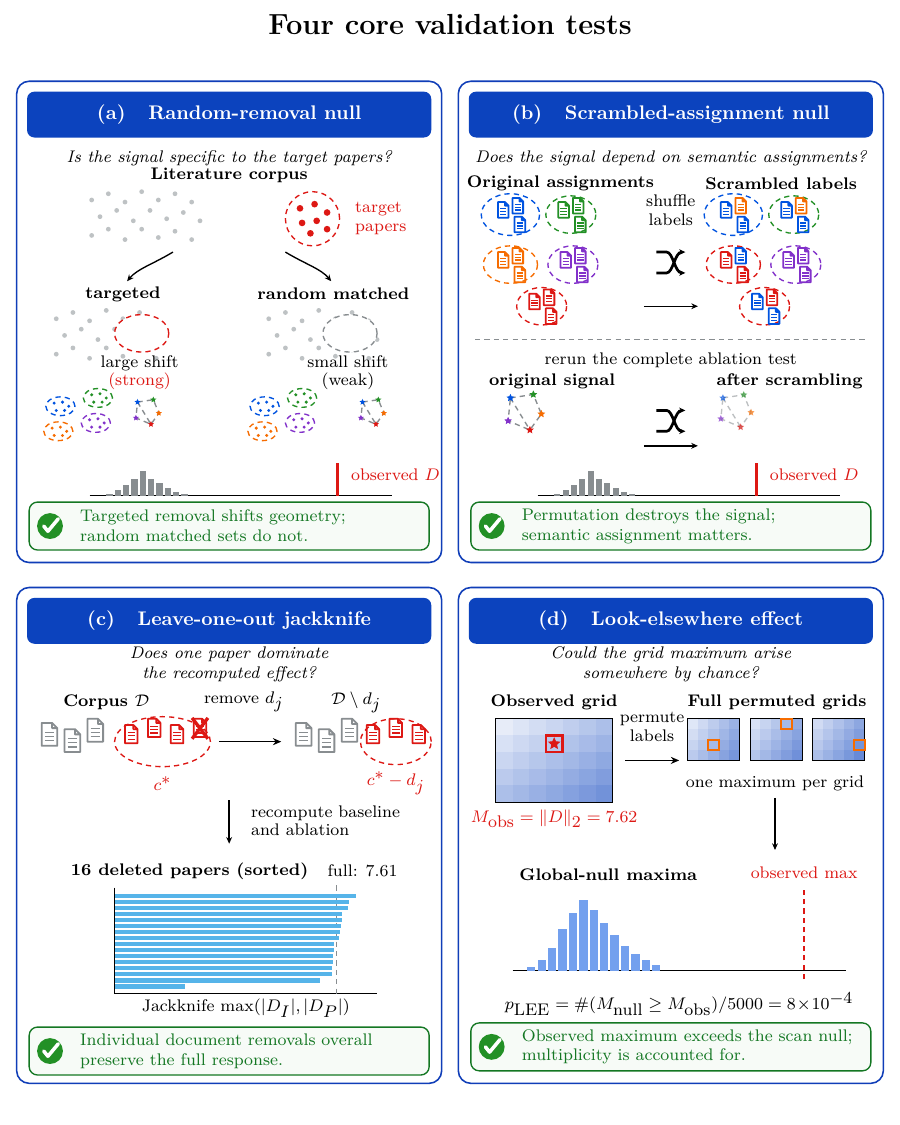}
\caption{Statistical validation suite used to evaluate the counterfactual framework. (a)~Random-removal null (targeted vs.\ random paper removal). (b)~Scrambled-assignment null (permuted concept labels). (c)~Leave-one-out stability (single-paper jackknife) and (d) Look-elsewhere effect.}
\label{fig:validation_suite}
\end{figure*}

The metric $D$ introduced in~\cref{sec:ablation} is intended to identify genuine conceptual reorganizations rather than artifacts arising from corpus composition, document assignment, or individual influential publications. We therefore evaluate every detected signal using a complementary suite of statistical validation tests designed to probe these potential sources of bias. The overall validation strategy is summarized in~\cref{fig:validation_suite}.

Four complementary tests are performed. The \emph{random-removal null} compares the observed geometric perturbation with the distribution obtained by removing temporally matched random sets of papers, testing whether the detected signal is specific to the target concept rather than a consequence of removing an arbitrary subset of the corpus~\citep{good2000permutation}. The \emph{scrambled-assignment null} randomly permutes concept assignments while preserving the temporal structure of the corpus, testing whether the observed perturbation depends on the correct semantic assignment of documents to concepts rather than on the overall corpus geometry~\citep{subramanian2005gsea}. Third, a \emph{leave-one-out} jackknife evaluates the stability of the standardized response $D$ against the deletion of individual documents, quantifying how much of the measured effect survives the loss of any single paper~\citep{efron1983loo}. Fourth, the \emph{look-elsewhere test} assesses whether the largest response in the concept $\times$ pivot-year scan remains significant after accounting for the full scan. Robustness with respect to document assignment is addressed independently through the threshold-selection procedure discussed in Section~\ref{sec:threshold}.

The complete validation procedure is applied to every historical case study. Detailed results are presented for the special relativity benchmark in Section~\ref{sec:results}, while the corresponding analyses for the remaining case studies are summarized in Section~\ref{sec:crosscase}.

In addition, the framework is evaluated for robustness with respect to implementation choices, including document assignment and embedding representation. These complementary robustness studies are presented along with the historical validation in Sections~\ref{sec:results} and~\ref{sec:crosscase}.

\section{Results: special relativity case study}
\label{sec:results}

\subsection{Detection of conceptual reorganization}

We first apply the proposed framework to the special relativity corpus, which serves as the primary benchmark for illustrating the methodology. 

The special relativity corpus contains 2,314 documents. Applying the document-assignment procedure described in Section~\ref{sec:embeddings} yields 62 documents assigned across the ten candidate concepts after rejecting 1,384 documents through the per-concept similarity threshold and a further 868 through the confusion-aware margin criterion. These assigned documents define the concept geometry analyzed throughout the remainder of this section. Not every candidate concept receives assigned documents under the high-confidence assignment criteria. Four concepts (gravitation, radiation/quantum, mechanics/time, and instrumentation) receive no document assignments, while two concepts (aether optics and electrodynamics) are represented by a single assigned document. The absence of assigned documents reflects the conservative high-confidence assignment criteria rather than the historical importance of these concepts, while  based on a single assigned document should be interpreted with appropriate caution.

At the data-driven pivot $t^*=1902$ obtained from the concept $\times$ pivot-year analysis (\cref{sec:grid}), we compute the counterfactual standardized response for each of the ten candidate concepts, shown in \cref{fig:grid} and summarized in \cref{tab:ranking}. Special relativity gives $\max(|D_I|,|D_P|)=7.61$, six times the second-ranked concept. Removing the special relativity literature therefore perturbs the surrounding embedding geometry far more than removing any other concept in the corpus.

%This is shown in~\cref{fig:grid}. The resulting effect sizes are summarized in \cref{tab:ranking}. Special relativity produces by far the largest effect, with a total-inertia effect size of $D_I=7.61$, substantially exceeding the second-ranked concept (spectroscopy, $\max|D|=1.26$). The large separation indicates that removing the special relativity literature produces a uniquely strong reorganization of the surrounding embedding geometry, consistent with the emergence of a new conceptual organizing principle.

The ranking also illustrates the interpretability of $D$. Several historically established concepts, including thermodynamics and aether optics, exhibit negative values of the total-inertia response $D_I$, indicating that their geometric influence is concentrated before the pivot year. Removing these concepts therefore perturbs the earlier embedding geometry more strongly than the later one, consistent with their declining structural role following the emergence of special relativity.

\begin{table}[htb]
\centering
\caption{Counterfactual standardized responses for all candidate concepts evaluated at the data-driven pivot year $t^*=1902$. Concepts are ranked by $\max(|D_I|,|D_P|)$.}
\label{tab:ranking}
\resizebox{\columnwidth}{!}{%
\begin{tabular}{@{}clrrrr@{}}
\toprule
Rank & Concept & $D_I$ & $D_P$ & $\max|D|$ & Papers \\
\midrule
1 & Special relativity         & 7.61  & $-$0.36  & 7.61 & 16 \\
2 & Spectroscopy               & 1.26  & 0.82   & 1.26 & 9 \\
3 & Aether optics              & $-$0.78 & $-$0.60 & 0.78 & 1 \\
4 & Electrodynamics            & $-$0.78 & $-$0.15 & 0.78 & 1 \\
5 & Electron theory            & 0.71  & 0.15   & 0.71 & 6 \\
6 & Thermodynamics             & $-$0.67 & $-$0.49  & 0.67 & 29 \\
--- & Gravitation                & ---  & ---  & --- & 0 \\
--- & Radiation / quantum        & ---  & ---  & --- & 0 \\
--- & Mechanics / time           & ---  & ---  & --- & 0 \\
--- & Instrumentation           & ---  & ---  & --- & 0 \\
\bottomrule
\end{tabular}}
\end{table}

\begin{figure}[t]
\centering
\includegraphics[width=\columnwidth]{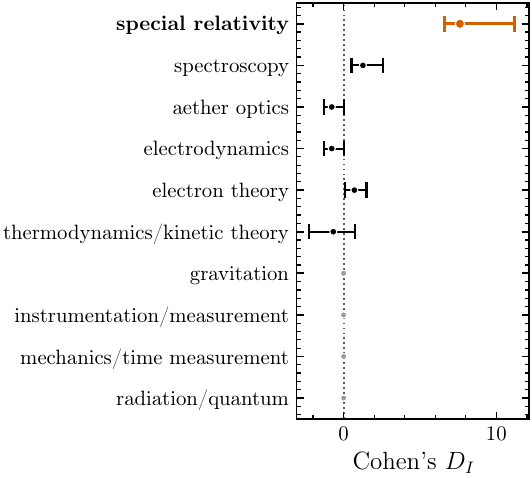}
  \caption{Ablation standardized responses $D_I$ at the pivot $t^{*}=1902$, for all ten concepts in the special relativity panel, ordered by $|D_I|$. For each concept, the point estimates and the corresponding 95\% bootstrap confidence intervals over 5{,}000 resamples are given. The four greyed concepts have no assigned documents before the pivot, so their perturbation is identically zero.}
\label{fig:ci}
\end{figure}

To quantify the stability of the observed effect, we estimate bootstrap confidence intervals by resampling the yearly ablation perturbations with replacement, separately within the pre-pivot and post-pivot periods, and recomputing the complete ranking for each resample. Bootstrap confidence intervals are shown in \cref{fig:ci}. Special relativity remains the highest-ranked concept in more than $99.9\%$ of bootstrap realizations, with a 95\% confidence interval of $D_I\in[6.58,\,11.19]$. Notably, the lower bound of this interval exceeds the point estimate of every other concept, demonstrating that the observed separation is not driven by statistical fluctuations or a small number of yearly observations.

The magnitude of the observed effect, however, is not by itself sufficient to establish that it represents a genuine conceptual reorganization. We therefore next examine the statistical significance and robustness of the special relativity signal using the validation framework introduced in Section~\ref{sec:validation}.

\subsection{Statistical validation of the special relativity signal}

As already explained, the validation suite tests four complementary hypotheses: whether the observed signal could arise from random document removal, whether it depends on the correct semantic assignment of documents to concepts, whether it is driven by a small number of influential papers, and whether the concept $\times$ pivot-year scan inflates its statistical significance.

\paragraph{Random-removal null.}

\begin{figure}[t]
\centering
\includegraphics[width=\columnwidth]{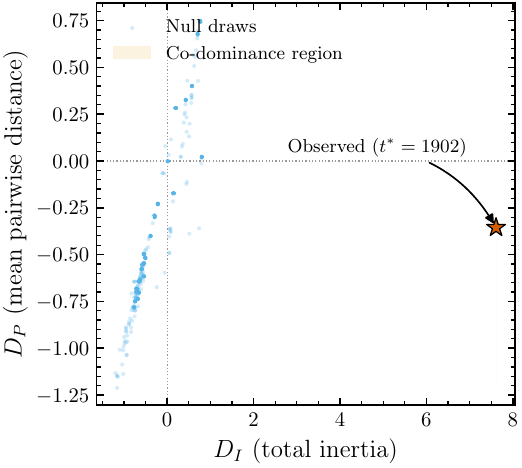}
\caption{Random-removal null on the joint $(D_I, D_P)$ plane for special relativity (5{,}000 draws of 16 temporally-matched random-paper sets, blue). The observed point at $t^{*}=1902$ is denoted with a red star. }
\label{fig:null}
\end{figure}

To test whether the observed perturbation could arise simply from removing an arbitrary subset of the corpus, we compare the observed effect with a null distribution obtained by 5,000 permutations of removing 16 temporally matched random papers, corresponding to the size of the special relativity cluster. This is shown in~\cref{fig:null}. The observed point $(D_I,D_P)=(7.61,-0.36)$ lies well outside the resulting null distribution on the joint $(D_I,D_P)$ plane ($p_{\rm co}=2.0\times10^{-4}$), with none of the 5,000 random realizations co-dominating the observed signal. This demonstrates that the measured perturbation is specific to the special relativity documents rather than a generic consequence of removing a similarly sized subset of the corpus.

\paragraph{Scrambled-assignment null.}

\begin{figure}[t]
\centering
\includegraphics[width=\columnwidth]{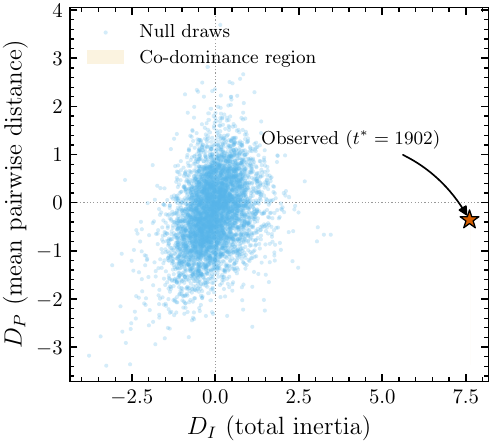}
\caption{Scrambled-assignment null on the joint $(D_I, D_P)$ plane for special relativity (5{,}000 label permutations, blue). The observed point at $t^{*}=1902$ is likewise denoted with a red star. }
\label{fig:scrambled}
\end{figure}

To test whether the observed signal depends on the specific semantic assignment of documents to concepts rather than simply removing an equally sized subset of papers, we construct a null distribution by randomly permuting concept labels across all assigned documents (5,000 permutations) while preserving the temporal structure of the corpus and recomputing the complete geometry for each realization. This is shown in~\cref{fig:scrambled}. The observed point again lies well outside the resulting null distribution on the joint $(D_I,D_P)$ plane ($p_{\rm co}=2.0\times10^{-4}$). This demonstrates that the detected perturbation depends on the specific documents associated with the special relativity concept rather than simply on the number of documents removed.

\paragraph{Leave-one-out stability.}

\begin{figure}[t]
\centering
\includegraphics[width=\columnwidth]{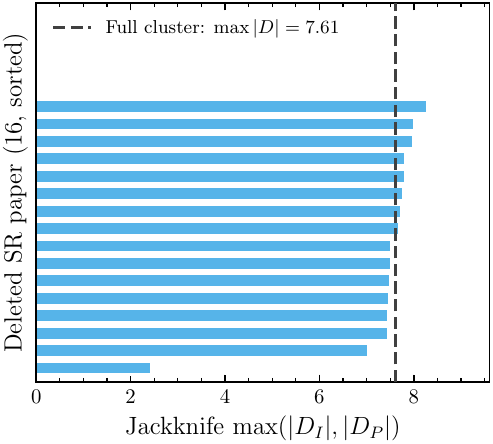}
\caption{Leave-one-out jackknife for special relativity. Each bar deletes one paper from the corpus and recomputes both the baseline and the ablated geometry over the remaining documents. The corresponding bar length shows the standardized response the analysis would have reported had that paper never been collected. The dashed line marks the full-concept value $\max(|D_I|,|D_P|)=7.61$ at $t^*=1902$.}
\label{fig:loo}
\end{figure}

The third test examines whether the observed geometric perturbation depends
disproportionately on any single document within the special relativity
concept. This is a robustness test of the framework rather
than a statement about the historical role of individual publications. We
delete one special relativity paper at a time from the corpus entirely,
recompute both the unablated baseline and the ablated geometry over the
remaining documents, and evaluate the standardized response at the data-driven pivot
$t^*=1902$. The results are shown in~\cref{fig:loo}. The complete concept
yields $\max(|D_I|,|D_P|)=7.61$, and the jackknife distribution has median
$7.61$, so for fifteen of the sixteen assigned papers the deletion changes
the measured effect by less than ten percent. One document, Cohn's 1904
alternative electrodynamics of moving systems, reduces the effect to
$2.43$, or $32\%$ of the full value; every other single deletion leaves it
essentially unchanged. The effect is therefore not an artifact of one
assigned document, although the concept is not entirely insensitive to its
composition.

\paragraph{Look-elsewhere effect.}

Finally, because the pivot year is determined through a scan over candidate concepts and years, the observed maximum must be assessed against the corresponding look-elsewhere effect~\cite{gross2010trial}. This is because the concept $\times$ pivot-year grid scans $10 \times 11 = 110$ cells and reports the global maximum, with a plausible concern being that with 110 opportunities to find a large $D$ by chance, the probability of observing a spurious maximum is inflated relative to a single hypothesis test. We estimate this by randomly permuting concept labels 5,000 times, as in the scrambled-assignment null, recomputing the complete concept $\times$ pivot-year grid for each permutation, and recording the maximum joint standardized response $\Dnorm=\sqrt{D_I^2+D_P^2}$ across all grid cells.

\begin{figure}[t]
\centering
\includegraphics[width=\columnwidth]{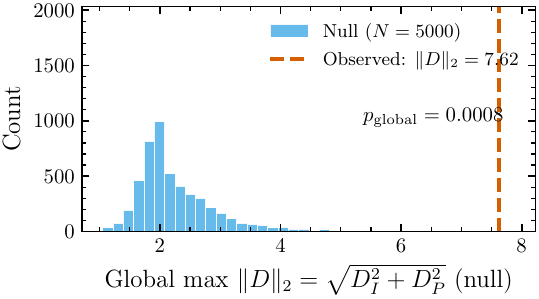}
\caption{Null distribution of the global maximum of the 2D magnitude $\Dnorm = \sqrt{D_I^2 + D_P^2}$ observed across the full $10 \times 11$ grid when concept labels are randomly permuted ($N = 5000$). The red dashed line corresponds to the observed SR signal ($\Dnorm = 7.62$).}
\label{fig:lee}
\end{figure}

This produces a null distribution of ``the largest joint signal one would
observe anywhere in the grid by chance'', and is shown for the special relativity case in~\cref{fig:lee}. The global $p$-value is then defined as the fraction of permutations whose maximum equals or exceeds the observed maximum. For the special relativity case, this procedure yields $p_{\rm LEE}=8 \cdot 10^{-4}$. Thus, the observed signal remains significant even after accounting for the multiplicity introduced by the concept $\times$ pivot-year scan.

These validation tests demonstrate that the special relativity signal is statistically significant, concept-specific, and robust against individual influential publications. We next examine the robustness of the framework with respect to the document-assignment procedure.

\subsection{Robustness of the document assignment}
\label{sec:threshold}

  The document-assignment procedure introduced in~\cref{sec:embeddings} requires an operating point controlling the balance between assignment efficiency and semantic contamination.
  Lower thresholds admit more documents at the expense of increased ambiguity between neighboring concepts, whereas higher thresholds improve semantic purity while discarding genuinely
  relevant documents.

  For a document $d$ with cosine similarities $\{s_d(c)\}_{c \in \mathcal{C}}$ to the concept anchors, the assignment margin is the gap between its best and second-best concept,
  \begin{equation}
  m_d = \max_{c \in \mathcal{C}} s_d(c) \; - \; \max_{c \neq c^{*}(d)} s_d(c),
  \qquad c^{*}(d) = \argmax_{c \in \mathcal{C}} s_d(c),
  \label{eq:margin}
  \end{equation}
  so a small $m_d$ marks a document lying close to the boundary between two concepts.

  The simplest criterion admits a document when $m_d$ exceeds a single threshold applied uniformly to all concepts. Such a threshold treats all concepts identically, ignoring that some
  concepts have close neighbors in embedding space, and so require a larger margin to separate, while others are isolated and admit a smaller margin without contamination. The framework
  therefore adopts a confusion-aware margin:

  \begin{equation}
  \tau(c)=f_{\rm knee}\left(1-s_{\rm nn}(c)\right),
  \label{eq:confusion}
  \end{equation}
  where $s_{\rm nn}(c)=\max_{c'\neq c}\hat{v}_{c}\cdot\hat{v}_{c'}$ is the cosine similarity between the unit-normalized key-idea anchor of concept $c$ and that of its nearest
  neighboring concept, so that $1-s_{\rm nn}(c)$ measures the angular separation between a concept and its closest competitor.

  A document is admitted to concept $c$ when both conditions hold: its margin satisfies $m_d \ge \tau(c)$, and its similarity to the concept anchor exceeds a per-concept baseline set at
  the sixtieth percentile of the similarity distribution of all documents whose nearest anchor is $c$. The percentile is fixed a priori and is not tuned. Documents failing either
  condition are labeled ``no match'' and excluded from the geometric analysis.

  The operating fraction $f_{\rm knee}$ is determined automatically using the Kneedle algorithm~\citep{satopaa2011kneedle}, which identifies the knee of the $\max|D|$ response curve
  obtained by scanning candidate operating points. The scan is performed on a grid of candidate operating fractions spanning $f\in[0.05,\,0.50]$, identical across encoders within a given case, and evaluated at the pivot year of the primary model, so that the threshold of every encoder in a given case is calibrated at a common reference year. The pivot year at which each encoder is subsequently evaluated is determined separately, by the procedure of~\cref{sec:modelrobustness}.

  The operating fraction is selected on the target concept's response curve.
  The selected fraction for every case study is reported in \cref{tab:crosscase}, and the statistical tests of \cref{sec:validation} are evaluated at the fixed operating point,
  where they assess the significance of the perturbation magnitude relative to the
  null distributions constructed at that same point.

  For the special relativity benchmark, the target retains the highest rank over a plateau around the selected operating point. The response curve exhibits its knee at $f_{\rm
  knee}=0.25$ ($D_I=7.61$), while special relativity remains the highest-ranked concept throughout the interval $f\in[0.22,\,0.30]$. The effect decreases when the threshold becomes
  sufficiently restrictive to over-prune the assigned documents or sufficiently permissive to introduce contamination from neighboring concepts. These results indicate that the
  identification of special relativity as the dominant concept is stable against the precise choice of assignment threshold, although the magnitude of the effect varies across the
  plateau and decreases sharply once the threshold prunes the assigned set to a few documents.

  The robustness of the assignment procedure across all historical case studies is examined in Section~\ref{sec:crosscase}.

\section{Historical validation across scientific revolutions}
\label{sec:crosscase}

The special relativity case study demonstrates that counterfactual ablation can detect a well-established conceptual reorganization and that the resulting signal is statistically robust. An important question, however, is whether this behavior is specific to a single historical example or reflects a more general property of conceptual change.

To investigate this question, we apply the same framework to four additional historical case studies: G\"odel's incompleteness theorems, the Higgs mechanism, the deep learning revolution, and the attention mechanism /
transformer architecture. Together, these examples encompass conceptual reorganizations spanning different scientific disciplines, historical contexts, and patterns of development, providing a broader test of the proposed framework.

Each corpus was assembled from Crossref~\citep{hendricks2020crossref}, zbMATH Open~\citep{schubotz2021zbmath}, and arXiv~\citep{ginsparg1994first}, embedded with e5-base, and analyzed using the same pipeline introduced in Section~\ref{sec:method}, including document embedding, concept assignment, counterfactual ablation, and statistical validation. Only corpus-specific inputs (document collections and concept definitions) differ between cases. Per-case corpus statistics, assignment rates, and notes on boundary contamination are summarized in \cref{tab:corpora}.

\begin{table*}[htb]
\centering
\caption{Cross-case corpus statistics.
         $r$ is the rolling-window radius in years (each window spans
         year $\pm r$); smaller corpora use $r=2$ to accumulate
         sufficient papers per window, while larger corpora use $r=1$
         for finer temporal resolution.
         Target = number of corpus documents assigned to the target
         concept; Target~\% = target papers as a fraction of total
         corpus documents.
         Margin-excl.\ = fraction of all corpus documents that clear the
         similarity floor but are nonetheless excluded
         because no concept wins the confusion-aware margin. Document counts refer to the analysis window shown; for deep learning the underlying corpus extends to 2020, and the additional documents are excluded from the geometry so that the case remains disjoint from the transformer era examined in the attention case.}
\label{tab:corpora}
%\resizebox{\columnwidth}{!}{%
\begin{tabular}{@{}lccccccc@{}}
\toprule
Case study & Documents & Time span & Assigned & Target & Target \% & Margin-excl. & $r$ \\
\midrule
Special relativity    & 2,314  & 1880--1920 & 62 (3\%)     & 16    & 0.7\% & 37.5\% & 2 \\
G\"odel incomp.       & 599    & 1900--1970 & 101 (17\%)    & 34    & 5.7\% & 23.4\% & 2 \\
Deep learning         & 16,152 & 2005--2018 & 4,053 (25\%) & 1,044 & 6.5\% & 14.9\% & 1 \\
Higgs mechanism       & 27,591 & 1955--1980 & 746 (3\%)    & 108   & 0.4\% & 37.3\% & 2 \\
Attention mech.       & 37,598 & 2005--2023 & 2,715 (7\%)  & 63    & 0.2\% & 32.8\% & 1 \\
\bottomrule
\end{tabular}
%}
\end{table*}

We first examine whether the standardized response consistently identifies the target conceptual reorganization across these independent historical case studies. We then evaluate whether the corresponding statistical and methodological validation tests exhibit the same behavior observed for the special relativity benchmark.

\subsection{Generalization across historical case studies}

%We now investigate whether the counterfactual observable generalizes beyond the special relativity benchmark. 

\begin{figure*}
\centering
\includegraphics[width=\textwidth]{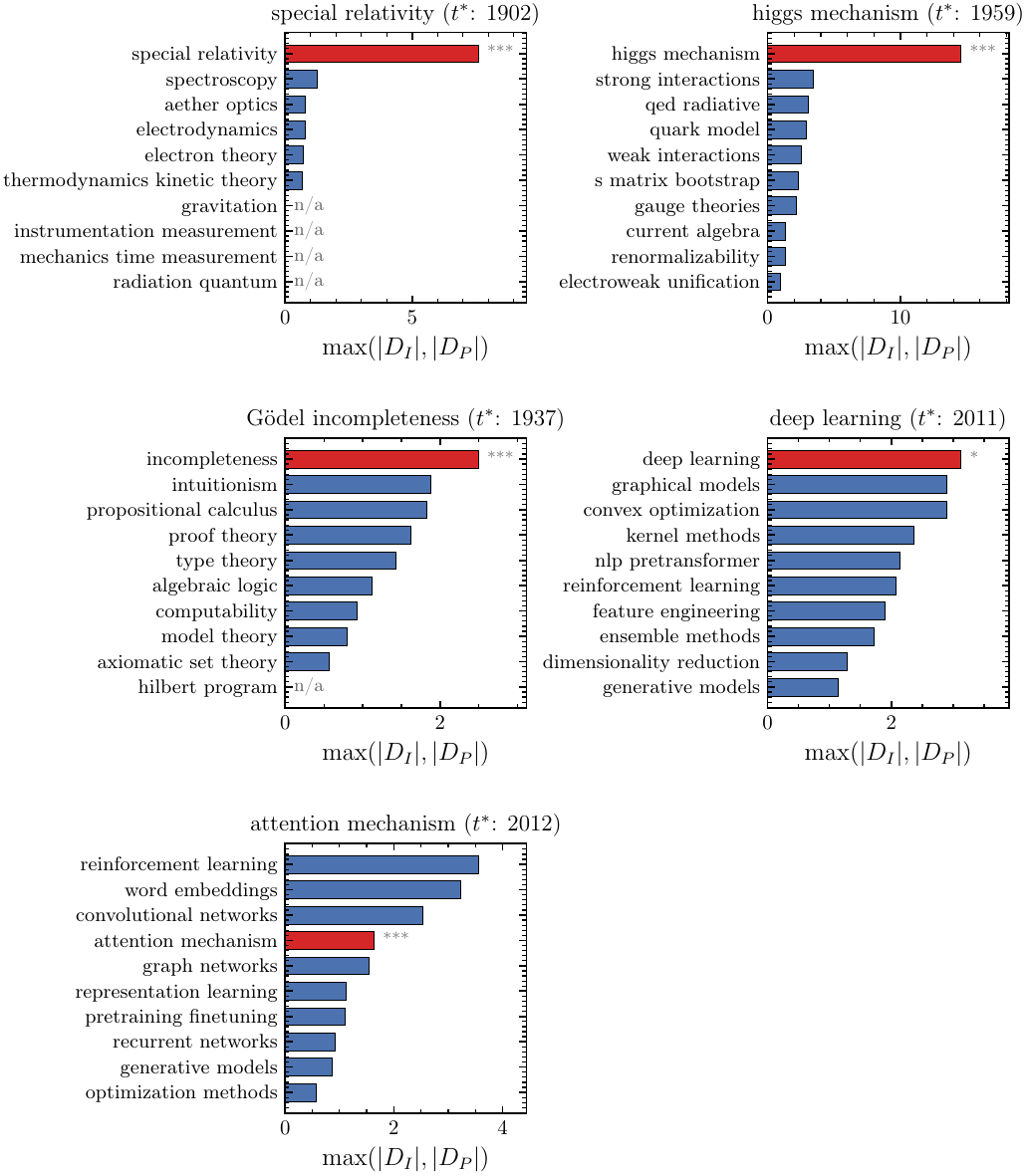}
\caption{Counterfactual standardized-response rankings across the five historical case studies:
         (a)~special relativity, (b)~Higgs mechanism,
         (c)~G\"odel incompleteness, (d)~deep learning,
         (e)~attention mechanism.  Red bars mark the target concept;
         blue bars are context concepts.  Stars indicate statistical
         significance ($^{*}p<0.05$, $^{**}p<0.01$, $^{***}p<0.001$)
         from the random-removal null.  Zero-height bars indicate
         concepts with no papers assigned under the confusion-aware
         margin; no ablation effect is computable for these concepts.}
\label{fig:crosscase}
\end{figure*}

Figure~\ref{fig:crosscase} and Table~\ref{tab:crosscase} summarize the Counterfactual standardized responses evaluated at the data-driven pivot year identified for each historical case study. In four of the five cases, the target concept produces the largest counterfactual geometric perturbation among the candidate concepts, indicating that the proposed metric consistently identifies the principal conceptual reorganization. The attention mechanism constitutes the only exception, ranking fourth within the broader and rapidly evolving deep-learning literature.

\begin{figure*}
\centering
\includegraphics[width=\textwidth]{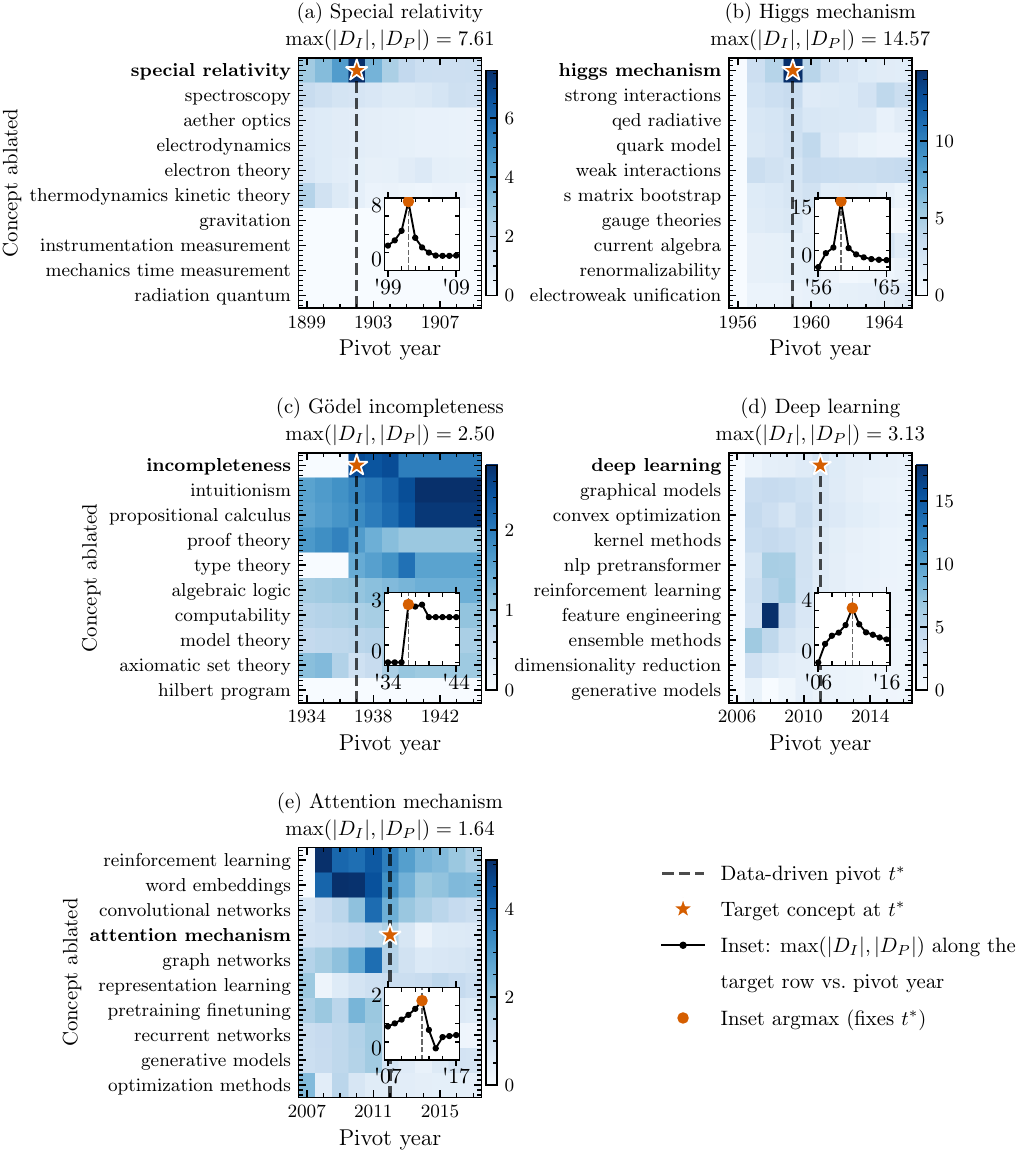}
\caption{Concept $\times$ pivot-year grids for the five historical case studies:
         (a)~special relativity, (b)~Higgs mechanism,
         (c)~G\"odel incompleteness, (d)~deep learning,
         (e)~attention mechanism.
         Each cell reports $\max(|D_I|,|D_P|)$ for ablating the row's concept
         and splitting pre- versus post-pivot at the column's year, the same
         ranking statistic listed in \cref{tab:crosscase}.
         The dashed line marks the data-driven pivot $t^{*}$ and the star marks
         the target concept at $t^{*}$.
         Rows are ordered by their value at $t^{*}$, so the ranking reads from
         top to bottom along the dashed line, and the target concept's label is
         set in bold. Color spans each panel's full grid, so cells at other years may exceed the starred cell without affecting the reported ranking, which compares concepts within the $t^{*}$ column only. The inset in each panel shows $\max(|D_I|,|D_P|)$ along the target concept's row against pivot year, over the same year range as the panel axis beneath it, with the endpoints of that range labelled by their final two digits; the orange marker is the argmax that defines $t^{*}$.}
\label{fig:crosscase_grid}
\end{figure*}

The full concept $\times$ pivot-year grids underlying these rankings are shown in \cref{fig:crosscase_grid}, which displays $\max(|D_I|,|D_P|)$ for every combination of ablated concept and candidate pivot year in all five case studies. Neither geometry observable ranks the target concept first in every case, so the two are combined through the same statistic used for the rankings in \cref{tab:crosscase} rather than selected case by case.

The two observables divide the five cases between them. Special relativity and the attention mechanism are dominated by the total-inertia response, whose positive sign means the target contributes more to the spread of the concept space after the pivot than before. Gödel incompleteness, the Higgs mechanism, and deep learning are dominated instead by the pairwise-distance response, whose sign measures whether the target became more or less distinguishable from the concepts around it: removing a concept lowers the mean pairwise distance when that concept sits farther from the others than they sit from each other, so a positive value means the target grew more distinct from its context across the pivot and a negative value that it grew less distinct. Gödel incompleteness is positive on this observable, and the Higgs mechanism and deep learning are negative, indicating targets that became progressively harder to separate from the surrounding literature as that literature expanded around them.

\begin{table*}
\centering
\small
\caption{Counterfactual standardized responses, confusion fractions, and null-test
significance for the five historical case studies.
$t^{*}$ denotes the target concept's argmax pivot, while $f_{\rm knee}$
is the per-case confusion fraction identified under e5-base.
$D_I$ and $D_P$ are Cohen's $D$ evaluated on the total-inertia and
mean-pairwise-distance geometry observables at $t^{*}$. Random removal and scrambled assignment report the co-dominance tail
$p_{\rm co}$,~\cref{eq:codominance},
on the $(D_I,D_P)$ plane using 5{,}000 permutations each.
The look-elsewhere effect (LEE) reports the global tail $p$-value of the
$\Dnorm$ grid maximum,
$\sqrt{D_I^2 + D_P^2}$, over all concept $\times$ pivot-year cells, also using
5{,}000 permutations.
Papers denotes the number of corpus documents assigned to the target concept.}
\label{tab:crosscase}

\begin{tabular}{@{}lrrrrcccr@{}}
\toprule
Case study
& $t^{*}$
& $f_{\rm knee}$
& $D_I$
& $D_P$
& \multicolumn{2}{c}{$p_{\rm co}$}
& LEE $p$
& Papers \\
\cmidrule(lr){6-7}
&
&
&
&
&
Random removal
& Scrambled assign.
&
& \\
\midrule

Special relativity
& 1902
& 0.25
& +7.61
& $-$0.36
& $2.0\times10^{-4}$ 
& $2.0\times10^{-4}$
& $8.0\times10^{-4}$
& 16 \\

Higgs mechanism
& 1959
& 0.30
& $-$3.25
& $-$14.57
& $2.0\times10^{-4}$
& $2.0\times10^{-4}$
& $2.0\times10^{-2}$
& 108 \\

G\"odel incompleteness
& 1937
& 0.10
& +0.50
& +2.50
& $2.0\times10^{-4}$
& $6.0\times10^{-4}$
& $1.5\times10^{-1}$
& 34 \\

Deep learning
& 2011
& 0.12
& $-$1.55
& $-$3.13
& $2.3\times10^{-2}$
& $2.0\times10^{-4}$
& $8.8\times10^{-2}$
& 1{,}044 \\

Attention mechanism
& 2012
& 0.25
& +1.64
& +0.90
& $8.0\times10^{-4}$
& $1.6\times10^{-2}$
& $9.7\times10^{-1}$
& 63 \\

\bottomrule
\end{tabular}
\end{table*}

\paragraph{Higgs mechanism (pivot 1959).}

The Higgs mechanism attains the largest nominal standardized response of all case studies, ranking first with $\max|D|=14.57$ from a cluster of 108 assigned papers. This magnitude is an artifact of pre-pivot sparsity. The data-driven pivot leaves only four pre-pivot yearly observations, and a single assigned document supplies the concept centroid in all four, so the pre-pivot perturbation is nearly constant, and the pooled variance entering the standardized response approaches zero. Removing that document reduces the response at 1959 to $1.24$ and moves the target from first to eighth among the ten candidate concepts, leaving the target row flat across the entire pivot range with no interior maximum. That document is a 1956 paper on cosmic time in general relativity, historically unrelated to electroweak symmetry breaking, admitted because the abstract description of a scalar field acquiring a background value and thereby reducing a symmetry fits both constructions. We therefore report the Higgs case as a diagnostic of unsupervised document assignment rather than as supporting evidence.

\paragraph{Gödel's incompleteness theorems (pivot 1937).}

Gödel's incompleteness theorems also rank first among the candidate concepts, with a maximum standardized response of $\max|D|=2.50$. The dominant contribution arises from the pairwise-distance response, consistent with a conceptual reorganization that primarily reshapes the relationships among neighboring areas of mathematical logic. The nearest competing concept is intuitionism, reflecting the close historical connection between these foundational developments.

\paragraph{Deep learning (pivot 2011).}

The deep learning revolution likewise ranks first, with $\max|D|=3.13$, but differs qualitatively from the preceding examples. The target concept encompasses a much larger literature (1,044 assigned documents), and the surrounding context concepts also exhibit substantial geometric perturbations. The resulting signal therefore reflects a broad paradigm shift distributed across many interacting research directions rather than a sharply localized conceptual transition. Consistent with this interpretation, the next most strongly perturbed concepts are graphical models and convex optimization ($D=2.89$ for both), major machine-learning paradigms that were progressively supplanted during the deep learning revolution.

\paragraph{Attention mechanism (pivot 2012).}

The attention mechanism provides a more demanding test of the framework by asking whether a localized conceptual innovation can be resolved within an ongoing scientific revolution. To study the attention mechanism, the machine-learning corpus was extended through 2023 to include the transformer era~\citep{vaswani2017attention}. 

Although the target concept, represented by 63 assigned papers, ranks fourth rather than first, it produces a statistically significant onset signal under both null tests, demonstrating that the standardized response retains sensitivity to individual conceptual developments embedded within a much broader transformation of the machine-learning landscape.

\paragraph{Comparison with surrounding concepts.}

\begin{figure}
\centering
\includegraphics[width=\columnwidth]{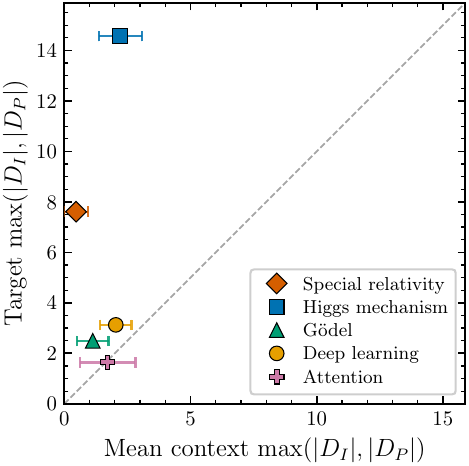}
  \caption{Target-concept standardized response $\max(|D_I|,|D_P|)$ at the
           data-driven pivot year $t^{*}$ versus the mean of the same statistic
           over the nine surrounding context concepts, for each historical case.
           Horizontal bars show one standard deviation of the context-concept
           responses. The dashed line is $y=x$: cases above it have the
           conceptual reorganization concentrated in the target concept, cases
           near or below it show diffuse reorganization.}
\label{fig:targetcontext}
\end{figure}

The relative importance of the target concept is summarized in~\cref{fig:targetcontext}, which compares the target standardized response with the mean standardized response of the surrounding context concepts for each historical case.   Special relativity exhibits pronounced target dominance, and G\"odel incompleteness a weaker but unambiguous one, indicating that the principal conceptual reorganization is strongly localized within the target concept. The Higgs mechanism shows the largest nominal separation of all cases, but for the reason given above this reflects the collapse of the pre-pivot variance rather than localization. Deep learning remains target-dominated but with a substantially smaller separation from the surrounding concepts, while the attention mechanism lies close to parity, reflecting its emergence within an already rapidly evolving research field.

\paragraph{Summary across historical case studies.}

Taken together, the five historical case studies show that the proposed framework can be applied across scientific disciplines and historical contexts. The framework consistently identifies conceptual reorganizations in four independent historical revolutions spanning physics, mathematics, and artificial intelligence, while also detecting a significant signal for the attention mechanism despite its emergence within a much broader ongoing transformation. These results demonstrate that counterfactual perturbations of embedding geometry provide a general quantitative measure of conceptual reorganization across diverse scientific domains and historical contexts.

%The observable identifies both sharply localized conceptual transitions and broader paradigm shifts, while remaining sensitive to conceptual developments embedded within ongoing scientific revolutions.

\subsection{Unified validation across historical case studies}

The statistical and methodological validation procedures introduced in Sections~\ref{sec:method} and~\ref{sec:results} were applied unchanged to all five historical case studies. These complementary tests assess the statistical significance, robustness, and temporal localization of the detected conceptual reorganizations.

The results of the statistical validation for the five test cases are summarized in~\cref{tab:crosscase}. Under both the random-removal and scrambled-assignment tests, all five historical case studies meet the significance criterion of $p<0.05$ defined in~\cref{sec:method}, confirming that the detected reorganizations depend on the specific conceptual organization of the corpus rather than on random document removal or arbitrary concept assignments. The leave-one-out analyses show that the measured responses are generally stable against individual-document removal, while also exposing an important failure mode. In the Higgs case, deletion of a single semantically misassigned pre-pivot document substantially reduces the standardized response, demonstrating that sparse pre-pivot assignments can generate artificially large values through variance collapse. This example illustrates the importance of the complementary validation procedures rather than weakening their role.

Applying the look-elsewhere test to the complete concept $\times$ pivot-year scan, the target concept supplies the global maximum for special relativity and the Higgs mechanism, whereas for Gödel's incompleteness theorems, deep learning, and the attention mechanism the largest perturbation in the grid is attained by a contextual concept and the global tail probability is not small. For the Higgs mechanism the global maximum inherits the pre-pivot variance collapse discussed above and does not carry independent evidential weight.

The methodological robustness studies likewise generalize across the historical case studies. Per-case, and across encoders, assignment thresholds derived using the confusion-aware margin span the range $f_{\rm knee}\in[0.05,\,0.45]$, reflecting differences in local concept-space density rather than arbitrary parameter choices. The resulting operating points and assignment statistics are summarized in \cref{tab:crosscase}.

Taken together, these validation studies show that the counterfactual framework can yield statistically significant and methodologically robust responses across diverse historical settings, while also identifying failure modes associated with sparse or contaminated document assignments.

\section{Robustness across embedding models}
\label{sec:modelrobustness}

The historical validation presented in Sections~\ref{sec:results} and~\ref{sec:crosscase} demonstrates that the proposed metric consistently detects conceptual reorganizations across multiple scientific domains. A remaining question is whether these conclusions depend on the particular embedding representation used to construct the document geometry. To investigate this, we repeat the complete analysis pipeline using the five embedding models introduced in~\cref{tab:embedding_models_main}. Because embedding models differ substantially in architecture, training objectives, and corpus coverage, agreement across models provides evidence that the observed geometric signatures reflect underlying conceptual organization rather than properties of a particular encoder.

\subsection{Dependence on embedding representation}

\begin{figure*}
\centering
\includegraphics[width=\textwidth]{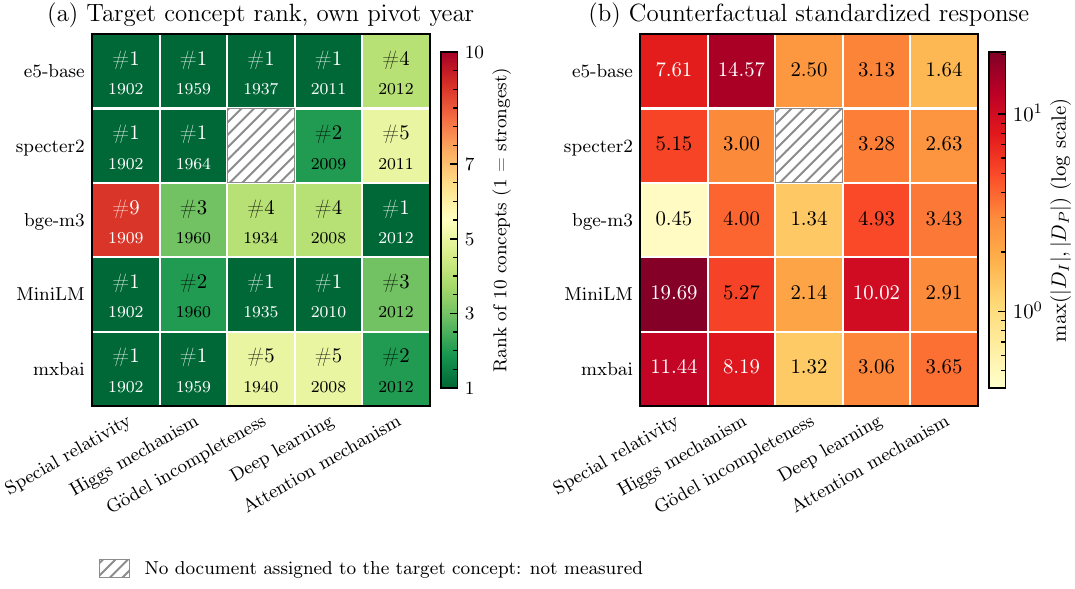}
%{outputs/cross_case_figures/fig_model_independence.pdf}
\caption{Embedding model independence across all five case studies. For each encoder, the pivot year is obtained by applying the target-row argmax rule of \cref{sec:grid} to that encoder's own concept $\times$ pivot-year grid. (a)~Target concept rank (green = high, red = low), with the selected pivot year printed beneath each rank. (b)~Standardized response $\max(|D_I|,|D_P|)$ evaluated at the same per-encoder pivot year. Hatched cells mark a case where the encoder assigns no document to the target concept, so neither the rank nor the standardized response is defined: SPECTER2 does not separate the Gödel concepts sufficiently for any document to satisfy both assignment criteria.}
\label{fig:modelindep}
\end{figure*}

Figure~\ref{fig:modelindep} and Table~\ref{tab:modelindep} summarize the target concept rank and counterfactual standardized response obtained for every combination of historical case study and embedding model. Each encoder is evaluated at the pivot year obtained by applying the same target-row argmax rule used for the primary model to that encoder's own concept $\times$ pivot-year grid. Consequently the comparison tests not only whether independent encoders agree on the relative standing of the target concept, but also whether they independently localize the same transition year. Because standardized-response magnitudes depend on the geometry of the embedding space, the principal quantity of interest is the ranking of the target concept rather than the absolute value of $|D|$.

\begin{table*}[t]
\centering
\caption{Target concept rank and standardized response across embedding models. The pivot year $t^{*}$ (shown per case) is derived from the concept $\times$ pivot-year grid (\cref{sec:grid}); each model is evaluated at its own target-row argmax; the selected year is listed beside each entry. Each model additionally uses its own independently optimized document-assignment threshold.}
\label{tab:modelindep}  
  \resizebox{\textwidth}{!}{%
  \begin{tabular}{@{}lccccccccccccccc@{}}
  \toprule
   & \multicolumn{3}{c}{Spec.\ Rel.} & \multicolumn{3}{c}{Higgs} & \multicolumn{3}{c}{G\"odel} & \multicolumn{3}{c}{Deep Learn.} & \multicolumn{3}{c}{Attention} \\
  \cmidrule(lr){2-4}\cmidrule(lr){5-7}\cmidrule(lr){8-10}\cmidrule(lr){11-13}\cmidrule(lr){14-16}
  Model & $t^{*}$ & Rank & $|D|$ & $t^{*}$ & Rank & $|D|$ & $t^{*}$ & Rank & $|D|$ & $t^{*}$ & Rank & $|D|$ & $t^{*}$ & Rank & $|D|$ \\
  \midrule
    e5-base  & 1902 & \#1/10 & 7.61  & 1959 & \#1/10 & 14.57 & 1937 & \#1/10 & 2.50 & 2011 & \#1/10 & 3.13  & 2012 & \#4/10 & 1.64 \\
    specter2 & 1902 & \#1/10 & 5.15  & 1964 & \#1/10 & 3.00  & --- & --- & --- & 2009 & \#2/10 & 3.28  & 2011 & \#5/10 & 2.63 \\
    bge-m3   & 1909 & \#9/10 & 0.45  & 1960 & \#3/10 & 4.00  & 1934 & \#4/10 & 1.34 & 2008 & \#4/10 & 4.93  & 2012 & \#1/10 & 3.43 \\
    MiniLM   & 1902 & \#1/10 & 19.69 & 1960 & \#2/10 & 5.27  & 1935 & \#1/10 & 2.14 & 2010 & \#1/10 & 10.02 & 2012 & \#3/10 & 2.91 \\
    mxbai    & 1902 & \#1/10 & 11.44 & 1959 & \#1/10 & 8.19  & 1940 & \#5/10 & 1.32 & 2008 & \#5/10 & 3.06  & 2012 & \#2/10 & 3.65 \\
  \bottomrule
  \end{tabular}}
\end{table*}

Across the five historical case studies, the principal conclusions remain largely stable across embedding models. For special relativity and the Higgs mechanism the target concept is ranked first by four and three of the five encoders, respectively, and for Gödel incompleteness, four of five place it among the five strongest concepts, while SPECTER2 assigns no document to the target and therefore yields no measurement. Deep learning and the attention mechanism have greater spread across encoders, with two and one encoders respectively, ranking the target first; in these two cases the variation affects the relative ordering of competing concepts rather than the existence of a geometric signal.

\paragraph{Sources of inter-model variation.}

Residual differences between embedding models arise primarily from differences in training objectives, corpus coverage, and representational capacity. Therefore, we assign each embedding model its own optimized confusion-aware assignment threshold, reflecting differences in embedding-space density. This is essential because a threshold appropriate for one encoder may over- or under-prune document assignments for another. Citation-prediction models (SPECTER2) compress semantically distinct but co-cited papers into overlapping regions, and in the most homogeneous corpus (the Gödel case) the confusion-aware threshold admits no document at all to the target concept, so that the cell reports a null result rather than a measured effect.  Low-dimensional distilled models (MiniLM) require tighter thresholds but produce strong signals once
properly calibrated.  Models trained on diverse web data (bge-m3)
require case-specific thresholds that differ substantially from those
optimized on the primary model.  Multilingual models (e5-base, mxbai)
preserve cross-lingual structure essential for historical corpora.

These results demonstrate that the proposed framework recovers the geometrical signal across encoders differing in architecture, training objective, and corpus coverage, but the concept ranking depends on the specific embedding representation used to construct the document geometry. Additionally, the localization of the transition year is not an artifact of the primary model: different encoders independently select pivot years that agree with the primary model exactly in about half of all cases and to within a few years in the large majority. The remaining differences between models may reflect differences in embedding objectives and semantic resolution. We therefore conclude that the quality of the semantic representation is itself an ingredient in measuring conceptual organization, and identifying which properties of an embedding space make conceptual reorganization measurable is a direction for future work.  

%SPECTER2 does not separate the Gödel concepts sufficiently for reliable assignment, illustrating that the success of the framework depends on the representational properties of the underlying embedding model.

\section{Discussion}
\label{sec:discussion}

The results presented in this work suggest that embedding geometry can function as a quantitative measure of conceptual reorganization. Across the historical case studies considered here, concentrated conceptual breakthroughs consistently produce stronger geometric signatures than broader, more diffuse transitions, indicating that major conceptual reorganizations leave measurable imprints within document embedding spaces. These findings naturally raise several interpretive questions. What do these geometric signatures represent? What assumptions underlie their interpretation? And how should their magnitude be understood in the context of scientific change? The following discussion addresses these questions and places the proposed framework within the broader landscape of computational approaches for studying the evolution of scientific knowledge.

\subsection{What the method detects}

Interpreting the results obtained requires distinguishing between the birth of a concept and its subsequent institutionalization within the scientific community. The proposed framework measures the latter: the stage at which a concept becomes a structural organizing principle around which a research community begins to organize its work. In this sense, the framework measures when a concept reshapes the conceptual organization of a field, rather than when it first enters the scientific record. In Kuhnian terms, a new ``paradigm" emerges~\citep{kuhn2012structure}.

The distinction is illustrated by the case of special relativity. Einstein's founding 1905 paper (\emph{``Zur Elektrodynamik bewegter K\"orper''}) is present in the corpus but is \emph{not} assigned to the special relativity concept cluster. Its nearest-concept similarity falls below the threshold, since the paper is framed kinematically rather than in the electrodynamic vocabulary that characterizes the assigned cluster, and the assignment rule therefore admits no concept for it. The assigned documents are instead dominated by the subsequent consolidation of special relativity as an established conceptual framework, including Minkowski's \emph{``Espace et temps''} (1909) and Einstein's own textbook \emph{Relativity: The Special and General Theory} (1916). No single one of them carries the signal: the drop-one jackknife of \cref{fig:loo} shows that fifteen of the sixteen assigned papers change the measured effect by less than ten percent.

This behavior follows naturally from interpreting embedding geometry as an observable of conceptual organization. A single paper, however revolutionary, does not immediately reorganize a field's embedding geometry. Reorganization occurs when a critical mass of researchers produces work organized around the new concept, creating a centroid whose position structurally depends on that concept. Consequently, the proposed framework cannot establish historical priority, i.e. who first introduced a concept, but it can identify when that concept became a structural organizing principle of the field.

An idea that is published but remains dormant for many years would likewise produce little or no geometric signature until it begins to reorganize the surrounding literature. From the perspective of the present framework, this is the expected behavior: the framework measures \emph{conceptual institutionalization} rather than invention itself. Accordingly, the geometric signatures reported throughout this paper should be interpreted as signatures of conceptual reorganization rather than signatures of individual acts of invention. Questions concerning originality or novelty require different methods, such as language-model-based analyses of individual publications~\citep{park2023papers}. These approaches are complementary rather than competing. Embedding geometry therefore measures when a concept becomes a structural organizing principle of a field, rather than when it first appears in the historical record.

%\subsection{Relation to prior work on the life of ideas}
\subsection{Relation to prior computational approaches}

\label{sec:relatedwork}

A variety of computational approaches have been developed to study how ideas emerge, evolve, and become established within scientific communities. \emph{Citation-based and bibliometric} methods characterize scientific change through the structure of the citation record itself, quantifying how combinations of prior work relate to subsequent impact~\citep{uzzi2013atypical} and how the disruptiveness of published work has evolved over time~\citep{park2023papers}. \emph{Topic-evolution} methods characterize how the prevalence and vocabulary of latent themes change over time. Dynamic topic models track topic-word distributions through time~\citep{blei2006dynamic,wang2008ctdtm}, while applications to scientific corpora recover disciplinary structure and the waxing and waning of research areas~\citep{griffiths2004finding,hall2008history}. \emph{Diachronic semantics} instead focuses on changes in the meaning of individual words using aligned historical or contextualized embeddings~\citep{hamilton2016diachronic,giulianelli2020lexical}, as surveyed by~\citet{kutuzov2018survey}. Finally, \emph{cultural evolution and idea diffusion} approaches quantify the adoption and spread of ideas through measures such as n-gram frequencies, citation patterns, adopter populations, and co-word networks. Examples include culturomics~\citep{michel2011culturomics}, epidemiological models of idea diffusion~\citep{bettencourt2006ideas}, phylomemetic analyses of scientific fields~\citep{chavalarias2013phylomemetic}, and studies of how ideas propagate and are reinterpreted across communities~\citep{keuchenius2021adoption,cheng2023diffuse}.

The proposed framework differs from these approaches in the quantity that it seeks to measure. Rather than quantifying the prevalence of a topic, the semantic evolution of a word, or the diffusion of an idea through citations or adoption, we quantify the structural role that a concept plays within the geometry of a document embedding space. The framework measures the geometric perturbation produced when the documents associated with a concept are counterfactually removed from the corpus. This provides a citation-independent, counterfactual framework for studying conceptual institutionalization.

Among existing approaches, the closest conceptual antecedents are the work of~\citet{cheng2023diffuse}, which distinguishes the introduction of an idea from its later integration into scientific practice, and~\citet{bettencourt2006ideas}, which treats community adoption as a measurable process. The methodological distinction of the present work lies in using embedding geometry and counterfactual ablation to quantify conceptual reorganization directly, without relying on citation networks or other external metadata.

\subsection{Retrospective validation and circularity}

A natural concern is that the framework is evaluated retrospectively using historically recognized conceptual reorganizations. We define concepts such as ``special relativity'', select a time window spanning the relevant historical period, and construct a corpus that includes the known innovation. Could the observed geometric signatures simply reflect assumptions built into the analysis rather than genuine conceptual reorganization?

Several features of the validation framework mitigate this concern. First, the concept $\times$ pivot-year grid evaluates \emph{all} candidate concepts across \emph{all} pivot years within the analysis window. If the results were merely a consequence of the predefined concept labels or historical window, one would expect multiple concept--year combinations to produce comparable signals. Instead, the special relativity $\times$ 1902 combination emerges as a sharp global maximum, substantially stronger than any non-special relativity concept--year pair.

Second, the validation suite tests whether the observed signal depends on the correct association between papers and concepts rather than on generic structural properties of the corpus or pivot-year grid.

Third, the framework does not require prior knowledge of which concept or historical year will produce the strongest signal. Given only a set of candidate concepts and an analysis window, the concept $\times$ pivot-year grid identifies which concept and which pivot year exhibit the largest geometric perturbation. In this sense, the historical examples serve as validation cases rather than predetermined outcomes.

A residual limitation nevertheless remains. The definition of corpus boundaries, candidate concepts, and temporal coverage necessarily involves analyst choices that may influence the quantitative results. This limitation is common to corpus-based approaches more generally and is not specific to the counterfactual ablation framework proposed here.

The purpose of the historical case studies is therefore not to rediscover known scientific revolutions, but to evaluate whether embedding geometry behaves as a meaningful observable of conceptual reorganization.

\subsection{Scope and interpretation}

Throughout this paper we have argued that embedding geometry provides a quantitative observable of conceptual reorganization. Interpreting this observable requires distinguishing it from several related but distinct notions, including novelty (the introduction of a new idea), conceptual reorganization (changes in the organization of a field), community institutionalization (the adoption of a concept by a research community), and scientific importance or impact. Although these phenomena are often correlated, they are not equivalent.

The counterfactual ablation framework introduced here measures one specific quantity: the geometric perturbation produced when the documents associated with a concept are removed from the embedding space. A large perturbation indicates that the concept has become structurally important in organizing the surrounding literature. Such signatures provide evidence that a concept has contributed to a measurable conceptual reorganization of the field. They do not, however, establish that the concept was historically original, scientifically correct, or ultimately influential in the long term. A concept that later proves to be incorrect could nevertheless reorganize the scientific literature during the period in which it is actively investigated.

Accordingly, the claims of the present work should be interpreted as identifying geometric signatures consistent with conceptual reorganization rather than providing a universal measure of innovation. The proposed framework quantifies structural change in the conceptual organization of scientific literature. Questions concerning originality, scientific value, historical priority, or long-term impact require complementary observables and lie beyond the scope of the present framework.

\subsection{Broader implications and future directions}

The present work establishes embedding geometry as a quantitative observable of conceptual reorganization. While the historical case studies considered here provide evidence that major conceptual reorganizations leave measurable geometric signatures, they also suggest a broader research direction. Having established embedding geometry as a quantitative observable of conceptual reorganization, it becomes possible to investigate how conceptual organization evolves over time.

Understanding the temporal evolution of embedding geometry represents a natural next step toward developing quantitative theories of knowledge evolution. Such studies may provide new insights into the dynamics of conceptual organization, the evolution of scientific disciplines, and, more generally, the processes through which knowledge develops across diverse domains. Beyond science, the same framework may prove useful for investigating conceptual change in other text-rich disciplines, including technology, philosophy, law, economics, and the humanities.

We leave these directions for future investigation. Here, our objective has been to establish the framework itself. We hope that the framework introduced here provides a foundation upon which future quantitative studies of conceptual organization and its evolution can be built.

\section{Conclusions}
\label{sec:conclusion}

This work introduces embedding geometry as a quantitative observable of conceptual reorganization and demonstrates that major conceptual reorganizations leave measurable geometric signatures within document embedding spaces. Across diverse historical examples, the proposed framework can identify principal conceptual reorganizations and distinguish localized conceptual breakthroughs from broader distributed paradigm shifts. Importantly, it does so without relying on citation networks, expert annotation, or other external metadata.

More broadly, our results suggest that the latent geometric structure learned by modern embedding models captures meaningful aspects of the conceptual organization encoded in scientific literature. Compared to current computational approaches to understanding scientific change, embedding geometry provides a complementary geometric perspective for studying how concepts organize, evolve, and influence the development of scientific fields.

Viewed from this perspective, modern embedding spaces can be regarded not only as semantic representations, but also as scientific objects of study whose geometry can be measured, perturbed, and statistically analyzed.

Treating embedding geometry as a quantitative scientific observable also introduces important limitations. The proposed framework measures the structural reorganization associated with the community uptake of ideas rather than their intrinsic originality or scientific value. Its validity depends on the quality of the underlying embedding representation and on the assignment of documents to concepts. The validation studies presented here indicate that the principal conclusions are robust across multiple embedding models, assignment procedures, and statistical tests, with one instructive exception: in the Higgs mechanism case a single misassigned document supplies the entire pre-pivot pool, and the reported standardized response is an artifact of the resulting variance collapse rather than a measure of conceptual reorganization.

A particularly promising next step is the development of \emph{time-bounded embedding} models, trained exclusively on literature available before a chosen historical date. Such models would eliminate information leakage from future developments, allowing conceptual reorganizations to be studied from the perspective of the scientific knowledge actually available at the time.  Beyond providing a more stringent historical validation, time-bounded embeddings would enable the evolution of embedding geometry itself to be investigated, opening the possibility of tracking how conceptual representations emerge, mature, and reorganize scientific fields. Such studies could provide the basis for quantitative models of knowledge evolution.

The broader ambition motivating this line of research is to develop a quantitative understanding of how knowledge evolves. Whether systems capable of recognizing the geometric signatures of past conceptual reorganizations can ultimately help identify emerging scientific directions, generate new hypotheses, or assist scientific discovery remains an open question. The retrospective evidence assembled here is a prerequisite for addressing that question rather than an answer to it: it establishes that embedding representations of scientific literature contain measurable geometric structure associated with conceptual reorganization, providing the foundation upon which future studies of knowledge evolution could build.

One particularly intriguing possibility is that such a framework could detect the ``quiet revolutions'' described by Kuhn~\citep{kuhn2012structure}: gradual conceptual reorganizations that unfold over decades before becoming widely recognized as transformative scientific advances. Understanding these dynamics may ultimately contribute to AI systems that not only analyze the history of science, but also assist scientists in understanding, and ultimately shaping, the future evolution of knowledge.
%assist scientists in exploring the future evolution of knowledge.

\section*{Acknowledgements}

We thank Malcolm Slaney for reviewing an earlier version of this manuscript and for valuable feedback. This work used the resources of the SLAC Shared Science Data Facility (S3DF) at SLAC National Accelerator Laboratory. SLAC is operated by Stanford University for the U.S. Department of Energy’s Office of Science.

\section*{Code and data availability}

The full analysis pipeline is released at \url{https://github.com/Mapping-Innovation-Lab/geometric-signatures}. Accompanying material is available on the paper website at \url{https://mapping-innovation-lab.github.io/geometric-signatures-companion-website/}.

\bibliographystyle{icml2026}
\bibliography{references}

\begin{thebibliography}{38}
\providecommand{\natexlab}[1]{#1}
\providecommand{\url}[1]{\texttt{#1}}
\expandafter\ifx\csname urlstyle\endcsname\relax
  \providecommand{\doi}[1]{doi: #1}\else
  \providecommand{\doi}{doi: \begingroup \urlstyle{rm}\Url}\fi

\bibitem[Bettencourt et~al.(2006)Bettencourt, Cintr{\'o}n-Arias, Kaiser, and Castillo-Ch{\'a}vez]{bettencourt2006ideas}
Bettencourt, L. M.~A., Cintr{\'o}n-Arias, A., Kaiser, D.~I., and Castillo-Ch{\'a}vez, C.
\newblock The power of a good idea: Quantitative modeling of the spread of ideas from epidemiological models.
\newblock \emph{Physica A: Statistical Mechanics and its Applications}, 364:\penalty0 513--536, 2006.

\bibitem[Blei \& Lafferty(2006)Blei and Lafferty]{blei2006dynamic}
Blei, D.~M. and Lafferty, J.~D.
\newblock Dynamic topic models.
\newblock In \emph{Proceedings of the 23rd International Conference on Machine Learning (ICML)}, pp.\  113--120, 2006.

\bibitem[Chavalarias \& Cointet(2013)Chavalarias and Cointet]{chavalarias2013phylomemetic}
Chavalarias, D. and Cointet, J.-P.
\newblock Phylomemetic patterns in science evolution---the rise and fall of scientific fields.
\newblock \emph{PLOS ONE}, 8\penalty0 (2):\penalty0 e54847, 2013.

\bibitem[Chen et~al.(2024)Chen, Xiao, Zhang, Luo, Lian, and Liu]{chen2024bge}
Chen, J., Xiao, S., Zhang, P., Luo, K., Lian, D., and Liu, Z.
\newblock {BGE M3-Embedding}: Multi-lingual, multi-functionality, multi-granularity text embeddings through self-knowledge distillation.
\newblock \emph{arXiv preprint arXiv:2402.03216}, 2024.

\bibitem[Cheng et~al.(2023)Cheng, Smith, Ren, Cao, Smith, and McFarland]{cheng2023diffuse}
Cheng, M., Smith, D.~S., Ren, X., Cao, H., Smith, S., and McFarland, D.~A.
\newblock How new ideas diffuse in science.
\newblock \emph{American Sociological Review}, 88\penalty0 (3):\penalty0 522--561, 2023.

\bibitem[Cohen(1988)]{cohen1988statistical}
Cohen, J.
\newblock \emph{Statistical Power Analysis for the Behavioral Sciences}.
\newblock Lawrence Erlbaum Associates, 2nd edition, 1988.

\bibitem[Efron(1983)]{efron1983loo}
Efron, B.
\newblock Estimating the error rate of a prediction rule: Improvement on cross-validation.
\newblock \emph{Journal of the American Statistical Association}, 78\penalty0 (382):\penalty0 316--331, 1983.

\bibitem[Einstein(1905)]{einstein1905electrodynamics}
Einstein, A.
\newblock Zur elektrodynamik bewegter k{\"o}rper.
\newblock \emph{Annalen der Physik}, 322\penalty0 (10):\penalty0 891--921, 1905.

\bibitem[Ethayarajh(2019)]{ethayarajh2019contextual}
Ethayarajh, K.
\newblock How contextual are contextualized word representations? {C}omparing the geometry of {BERT}, {ELMo}, and {GPT}-2 embeddings.
\newblock In \emph{Proceedings of the 2019 Conference on Empirical Methods in Natural Language Processing and the 9th International Joint Conference on Natural Language Processing}, pp.\  55--65, 2019.

\bibitem[Fortunato et~al.(2018)Fortunato, Bergstrom, B{\"o}rner, Evans, Helbing, Milojevi{\'c}, Petersen, Radicchi, Sinatra, Uzzi, et~al.]{fortunato2018science}
Fortunato, S., Bergstrom, C.~T., B{\"o}rner, K., Evans, J.~A., Helbing, D., Milojevi{\'c}, S., Petersen, A.~M., Radicchi, F., Sinatra, R., Uzzi, B., et~al.
\newblock Science of science.
\newblock \emph{Science}, 359\penalty0 (6379):\penalty0 eaao0185, 2018.

\bibitem[Ginsparg(1994)]{ginsparg1994first}
Ginsparg, P.
\newblock First steps towards electronic research communication.
\newblock \emph{Computers in Physics}, 8\penalty0 (4):\penalty0 390--396, 1994.

\bibitem[Giulianelli et~al.(2020)Giulianelli, Del~Tredici, and Fern{\'a}ndez]{giulianelli2020lexical}
Giulianelli, M., Del~Tredici, M., and Fern{\'a}ndez, R.
\newblock Analysing lexical semantic change with contextualised word representations.
\newblock In \emph{Proceedings of the 58th Annual Meeting of the Association for Computational Linguistics}, pp.\  3960--3973, 2020.

\bibitem[G{\"o}del(1931)]{godel1931undecidable}
G{\"o}del, K.
\newblock {\"U}ber formal unentscheidbare s{\"a}tze der {P}rincipia {M}athematica und verwandter systeme {I}.
\newblock \emph{Monatshefte f{\"u}r Mathematik und Physik}, 38:\penalty0 173--198, 1931.

\bibitem[Good(2000)]{good2000permutation}
Good, P.~I.
\newblock \emph{Permutation Tests: A Practical Guide to Resampling Methods for Testing Hypotheses}.
\newblock Springer, 2000.

\bibitem[Griffiths \& Steyvers(2004)Griffiths and Steyvers]{griffiths2004finding}
Griffiths, T.~L. and Steyvers, M.
\newblock Finding scientific topics.
\newblock \emph{Proceedings of the National Academy of Sciences}, 101\penalty0 (Suppl. 1):\penalty0 5228--5235, 2004.

\bibitem[Gross \& Vitells(2010)Gross and Vitells]{gross2010trial}
Gross, E. and Vitells, O.
\newblock Trial factors for the look elsewhere effect in high energy physics.
\newblock \emph{The European Physical Journal C}, 70:\penalty0 525--530, 2010.

\bibitem[Hall et~al.(2008)Hall, Jurafsky, and Manning]{hall2008history}
Hall, D., Jurafsky, D., and Manning, C.~D.
\newblock Studying the history of ideas using topic models.
\newblock In \emph{Proceedings of the 2008 Conference on Empirical Methods in Natural Language Processing (EMNLP)}, pp.\  363--371, 2008.

\bibitem[Hamilton et~al.(2016)Hamilton, Leskovec, and Jurafsky]{hamilton2016diachronic}
Hamilton, W.~L., Leskovec, J., and Jurafsky, D.
\newblock Diachronic word embeddings reveal statistical laws of semantic change.
\newblock In \emph{Proceedings of the 54th Annual Meeting of the Association for Computational Linguistics (Volume 1: Long Papers)}, pp.\  1489--1501, 2016.

\bibitem[Hendricks et~al.(2020)Hendricks, Tkaczyk, Lin, and Feeney]{hendricks2020crossref}
Hendricks, G., Tkaczyk, D., Lin, J., and Feeney, P.
\newblock Crossref: The sustainable source of community-owned scholarly metadata.
\newblock \emph{Quantitative Science Studies}, 1\penalty0 (1):\penalty0 414--427, 2020.

\bibitem[Higgs(1964)]{higgs1964broken}
Higgs, P.~W.
\newblock Broken symmetries and the masses of gauge bosons.
\newblock \emph{Physical Review Letters}, 13\penalty0 (16):\penalty0 508--509, 1964.

\bibitem[Keuchenius et~al.(2021)Keuchenius, T{\"o}rnberg, and Uitermark]{keuchenius2021adoption}
Keuchenius, A., T{\"o}rnberg, P., and Uitermark, J.
\newblock Adoption and adaptation: A computational case study of the spread of granovetter's weak ties hypothesis.
\newblock \emph{Social Networks}, 66:\penalty0 10--25, 2021.

\bibitem[Krizhevsky et~al.(2012)Krizhevsky, Sutskever, and Hinton]{krizhevsky2012imagenet}
Krizhevsky, A., Sutskever, I., and Hinton, G.~E.
\newblock {ImageNet} classification with deep convolutional neural networks.
\newblock In \emph{Advances in Neural Information Processing Systems}, volume~25, pp.\  1097--1105, 2012.

\bibitem[Kuhn(1962)]{kuhn1962structure}
Kuhn, T.~S.
\newblock \emph{The Structure of Scientific Revolutions}.
\newblock University of Chicago Press, Chicago, 1962.

\bibitem[Kuhn(2012)]{kuhn2012structure}
Kuhn, T.~S.
\newblock \emph{The Structure of Scientific Revolutions}.
\newblock University of Chicago Press, Chicago, 4 edition, 2012.
\newblock 50th Anniversary Edition, with an introductory essay by Ian Hacking.

\bibitem[Kutuzov et~al.(2018)Kutuzov, {\O}vrelid, Szymanski, and Velldal]{kutuzov2018survey}
Kutuzov, A., {\O}vrelid, L., Szymanski, T., and Velldal, E.
\newblock Diachronic word embeddings and semantic shifts: A survey.
\newblock In \emph{Proceedings of the 27th International Conference on Computational Linguistics (COLING)}, pp.\  1384--1397, 2018.

\bibitem[Lee et~al.(2024)Lee, Shakir, Koenig, and Lipp]{emde2024mxbai}
Lee, S., Shakir, A., Koenig, D., and Lipp, J.
\newblock Open source strikes bread - new fluffy embedding model.
\newblock \url{https://www.mixedbread.com/blog/mxbai-embed-large-v1}, March 2024.

\bibitem[Michel et~al.(2011)Michel, Shen, Aiden, Veres, Gray, {Google Books Team}, Pickett, Hoiberg, Clancy, Norvig, Orwant, Pinker, Nowak, and Aiden]{michel2011culturomics}
Michel, J.-B., Shen, Y.~K., Aiden, A.~P., Veres, A., Gray, M.~K., {Google Books Team}, Pickett, J.~P., Hoiberg, D., Clancy, D., Norvig, P., Orwant, J., Pinker, S., Nowak, M.~A., and Aiden, E.~L.
\newblock Quantitative analysis of culture using millions of digitized books.
\newblock \emph{Science}, 331\penalty0 (6014):\penalty0 176--182, 2011.

\bibitem[Park et~al.(2023)Park, Leahey, and Funk]{park2023papers}
Park, M., Leahey, E., and Funk, R.~J.
\newblock Papers and patents are becoming less disruptive over time.
\newblock \emph{Nature}, 613:\penalty0 138--144, 2023.

\bibitem[Reimers \& Gurevych(2019)Reimers and Gurevych]{reimers2019sentencebert}
Reimers, N. and Gurevych, I.
\newblock Sentence-{BERT}: Sentence embeddings using siamese {BERT}-networks.
\newblock In \emph{Proceedings of the 2019 Conference on Empirical Methods in Natural Language Processing and the 9th International Joint Conference on Natural Language Processing}, pp.\  3982--3992, 2019.

\bibitem[Satopaa et~al.(2011)Satopaa, Albrecht, Irwin, and Raghavan]{satopaa2011kneedle}
Satopaa, V., Albrecht, J., Irwin, D., and Raghavan, B.
\newblock Finding a ``kneedle'' in a haystack: Detecting knee points in system behavior.
\newblock In \emph{2011 31st International Conference on Distributed Computing Systems Workshops}, pp.\  166--171. IEEE, 2011.

\bibitem[Schubotz \& Teschke(2021)Schubotz and Teschke]{schubotz2021zbmath}
Schubotz, M. and Teschke, O.
\newblock {zbMATH Open}: Towards standardized machine interfaces to expose bibliographic metadata.
\newblock \emph{European Mathematical Society Magazine}, \penalty0 (119):\penalty0 50--53, 2021.

\bibitem[Singh et~al.(2023)Singh, D'Arcy, Cohan, Downey, and Feldman]{singh2023specter2}
Singh, A., D'Arcy, M., Cohan, A., Downey, D., and Feldman, S.
\newblock {SciRepEval}: A multi-format benchmark for scientific document representations.
\newblock In \emph{Proceedings of the 2023 Conference on Empirical Methods in Natural Language Processing}, 2023.

\bibitem[Subramanian et~al.(2005)Subramanian, Tamayo, et~al.]{subramanian2005gsea}
Subramanian, A., Tamayo, P., et~al.
\newblock Gene set enrichment analysis: a knowledge-based approach for interpreting genome-wide expression profiles.
\newblock \emph{Proceedings of the National Academy of Sciences}, 102\penalty0 (43):\penalty0 15545--15550, 2005.

\bibitem[Uzzi et~al.(2013)Uzzi, Mukherjee, Stringer, and Jones]{uzzi2013atypical}
Uzzi, B., Mukherjee, S., Stringer, M., and Jones, B.
\newblock Atypical combinations and scientific impact.
\newblock \emph{Science}, 342\penalty0 (6157):\penalty0 468--472, 2013.

\bibitem[Vaswani et~al.(2017)Vaswani, Shazeer, Parmar, Uszkoreit, Jones, Gomez, Kaiser, and Polosukhin]{vaswani2017attention}
Vaswani, A., Shazeer, N., Parmar, N., Uszkoreit, J., Jones, L., Gomez, A.~N., Kaiser, {\L}., and Polosukhin, I.
\newblock Attention is all you need.
\newblock In \emph{Advances in Neural Information Processing Systems}, volume~30, pp.\  5998--6008, 2017.

\bibitem[Wang et~al.(2008)Wang, Blei, and Heckerman]{wang2008ctdtm}
Wang, C., Blei, D., and Heckerman, D.
\newblock Continuous time dynamic topic models.
\newblock In \emph{Proceedings of the 24th Conference on Uncertainty in Artificial Intelligence (UAI)}, pp.\  579--586, 2008.

\bibitem[Wang et~al.(2024)Wang, Yang, Huang, Yang, Majumder, and Wei]{wang2024e5}
Wang, L., Yang, N., Huang, X., Yang, L., Majumder, R., and Wei, F.
\newblock Multilingual {E5} text embeddings: A technical report.
\newblock \emph{arXiv preprint arXiv:2402.05672}, 2024.

\bibitem[Wang et~al.(2020)Wang, Wei, Dong, Bao, Yang, and Zhou]{wang2020minilm}
Wang, W., Wei, F., Dong, L., Bao, H., Yang, N., and Zhou, M.
\newblock {MiniLM}: Deep self-attention distillation for task-agnostic compression of pre-trained transformers.
\newblock In \emph{Advances in Neural Information Processing Systems}, volume~33, pp.\  5776--5788, 2020.

\end{thebibliography}

%\appendix
% cleveref: label sectioning units defined below as "Appendix" (not "Section")
%\crefalias{section}{appendix}
%\crefalias{subsection}{appendix}
%\section*{Appendix}

\end{document}